\documentclass{emulateapj}
\usepackage{graphicx}
\usepackage{bm}

\newcommand{\Rta}{R_{\rm ta}}
\newcommand{\tta}{t_{\rm ta}}

\newcommand{\Kta}{K_{\rm ta}}
\newcommand{\tcol}{t_{\rm col}}
\newcommand{\Wv}{W_{\rm vir}}
\newcommand{\tv}{t_{\rm vir}}
\newcommand{\Kv}{K_{\rm vir}}
\newcommand{\Rv}{R_{\rm vir}}
\newcommand{\Deltav}{\Delta_{\rm vir}}
\newcommand{\alphata}{\alpha_{\rm ta}}
\newcommand{\betata}{\beta^0_{\rm ta}}
\newcommand{\gammata}{\gamma_{\rm ta}}
\newcommand{\Rmod}{R_{0,\rm TSC}}
\newcommand{\Rsim}{R_{0,\rm sim}}
\newcommand{\Rini}{R_{\rm ini}}
\newcommand{\alphai}{\alpha_{\rm ini}}
\newcommand{\betai}{\beta^0_{\rm ini}}
\newcommand{\gammai}{\gamma_{\rm ini}}
\newcommand{\deltai}{\delta_{\rm ini}}
\newcommand{\zta}{z_{\rm ta}}

\newcommand{\zini}{z_{\rm ini}}

\begin{document}

\title{Confrontation of Top-Hat Spherical Collapse against Dark Halos from Cosmological N-Body Simulations}

\author{
Daichi Suto \altaffilmark{1}, 
Tetsu Kitayama \altaffilmark{2}, 
Ken Osato \altaffilmark{1},
Shin Sasaki \altaffilmark{3},
Yasushi Suto \altaffilmark{1,4}
}

\email{daichi@utap.phys.s.u-tokyo.ac.jp}

\altaffiltext{1}{Department of Physics, The University of Tokyo, Tokyo 113-0033, Japan}
\altaffiltext{2}{Department of Physics, Toho University,  Funabashi, Chiba 274-8510, Japan}
\altaffiltext{3}{Department of Physics, Tokyo Metropolitan University, Hachioji, Tokyo 192-0397, Japan}
\altaffiltext{4}{Research Center for the Early Universe, School of Science, The University of Tokyo, Tokyo 113-0033, Japan}


\begin{abstract}
The top-hat spherical collapse model (TSC) is one of the most fundamental analytical frameworks to describe the non-linear growth of cosmic structure. TSC has motivated, and been widely applied in, various researches even in the current era of precision cosmology.  While numerous studies exist to examine its validity against numerical simulations in a statistical fashion, there are few analyses to compare the TSC dynamics in an individual object-wise basis, which is what we attempt in the present paper. We extract 100 halos at $z=0$ from a cosmological N-body simulation according to the conventional TSC criterion for the spherical over-density.  Then we trace back their spherical counter-parts at earlier epochs. Just prior to the turn-around epoch of the halos, their dynamics is well approximated by TSC, but their turn-around epochs are systematically delayed and the {\it virial} radii are larger by $\sim$ 20\% on average relative to the TSC predictions. We find that this systematic deviation is mainly ascribed to the non-uniformity/inhomogeneity of dark matter density profiles and the non-zero velocity dispersions, both of which are neglected in TSC. In particular, the inside-out-collapse and shell-crossing of dark matter halos play an important role in generating the significant velocity dispersion. The implications of the present result are briefly discussed.
\end{abstract}

\section{Introduction}

The top-hat spherical collapse \citep[TSC, hereafter;][]{Gunn72, Gunn77, Peebles80} is among the most fundamental analytical models that describe the non-linear cosmic structure formation. TSC, however, assumes a highly idealized system; a uniform, spherical and isolated object with no velocity dispersion until just before relaxation. In reality, the evolution of an actual object is not spherical nor isolated, and interacts with surrounding matter, but the TSC predictions are widely used in modern precision cosmology, including galaxy cluster scaling relations \citep[][and references therein]{Allen11}, halo identification in numerical simulations, and the Press-Schechter theory \citep{Press74} among others.

In particular, the halo identification combined with the Press-Schchter theory and TSC predictions provides the basis for determining cosmological parameters with the cluster abundance \citep[e.g.,][for recent studies]{Vikhlinin09, Rozo10, Reichardt13, Hasselfield13, Planck14}. Even though the halo abundance itself is computed from N-body simulations \citep{Jenkins01,Tinker08}, the definition of halos is indeed strongly motivated by the TSC predictions.

Several authors have attempted to improve TSC. A prime example is the self-similar spherical collapse model in the Einstein-de Sitter (EdS) universe, developed by \cite{Filmore84} and \cite{Bertschinger85}. This model adopts the radial the density profile of the initial fluctuations, which results in the velocity dispersion developing from the center to the outer region. \cite{Zukin10} have further derived self-similar solutions with tidal torque.

Another way of improvement is to abandon the spherical symmetry. \cite{White79} formulated the evolution of a tri-axial halo neglecting external tide, and \cite{Bond96} incorporated the effect of the external tide and the ``virialization'' of halos. The resulting ``ellipsoidal collapse model'' was applied by \cite{Sheth02} in improving the mass function of halos. Just like TSC, the ellipsoidal collapse model is based on many simplified assumptions concerning the halo evolution, so the validity of the model predictions are tested by several authors using cosmological simulations \cite[e.g.][]{Despali13, Despali14, Ludlow14, Borzyszkowski14}. 

\cite{Rossi11} have pointed out the disagreement between the prediction of the ellipsoidal collapse model and simulated halos; the ellipsoidal collapse model predicts that heavier objects are more spherical while in simulations heavier halos tend to have larger ellipticity \citep{Jing02, Despali14}. While the measurements of the ellipticity of galaxy clusters by X-ray \citep{Kawahara10} and weak lensing \citep{Oguri10} are consistent with the simulation results, the origin of the discrepancy is not yet clear.

Since it is difficult to handle the non-sphericity to model the halo evolution, we now return to TSC, and investigate how the TSC predictions hold when the idealized assumptions are broken. In fact, the validity of TSC predictions for an {\it individual} halo has not yet been adequately tested despite the fact that TSC is widespread as a {\it statistical} tool in modern cosmology. If we stack the initial conditions of those halos, the resulting average density profiles are spherically symmetric. Nevertheless the individual halo dynamics does not proceed in a spherically symmetric manner, and the final state should deviate from the TSC prediction. This trivial statement that the averaging and the dynamics do not commute is so obvious, but has never been examined carefully.

In this paper, we quantitatively compare the predictions of TSC with the evolution of 100 individual halos extracted from a cosmological N-body simulation, rather than considering their statistical average as has been performed so far.

The rest of this paper is organized as follows. In Section 2, we summarize the predictions of TSC, which will be confronted in the following sections. Section 3 describes our N-body simulation and how to trace back the simulated halos identified at present to the past. We there show the velocity dispersion of dark matter plays an important role, and investigate its evolution in the phase space. The evolution of the simulated halos is compared with TSC in Section 4. In Section 5, the main part of this paper, we examine how the velocity dispersion affects the ``virialization'' process of halos. Finally, Section 6 summarizes this paper.

\section{Top-hat spherical dust model}

This section reviews the basic assumptions and analytical predictions of TSC, for convenience in the following sections. First we describe the collapse model in the EdS universe, and then add the relevant modifications for the cosmological constant.

TSC assumes the following initial density profile:
\begin{eqnarray}
\rho_{\rm ini}\left\{
\begin{array}{ll}
\bar{\rho}_{\rm ini}(1+\deltai)
&;
\quad r\le \Rini\\
0
&;
\quad r>\Rini,
\end{array}
\right.
\end{eqnarray}
where $\bar{\rho}_{\rm ini}=(6\pi G t_{\rm ini}^2)^{-1}$ is the cosmic mean density at the initial time $t_{\rm ini}$, and $\deltai (>0)$ is the initial overdensity to the cosmic mean density. Throughout this paper, we refer to TSC for the uniform density ($\deltai=$ const.). Thus the total mass $M$ of the sphere is $4\pi \bar{\rho}_{\rm ini}(1+\deltai)\Rini^3/3$.

The time evolution of the outermost radius $R(t)$ of the sphere follows
\begin{equation}
\frac{d^2R}{dt^2}=-\frac{GM}{R^2},
\label{eomtsc}
\end{equation}
which has the cycloidal solution:
\begin{eqnarray}
\label{toftheta}
t &=& \frac{\tta}{\pi}(\theta-\sin\theta), \\
\label{Roftheta}
R &=&\frac{\Rta}{2}(1-\cos\theta).
\end{eqnarray}
It is conventional to normalize $t$ and $R$ by the values at the maximum expansion, or {\it turn-around} time, $\tta$ and turn-around radius $\Rta$, respectively, and they satisfy
\begin{equation}
\frac{\Rta^3}{\tta^2}\frac{\pi^2}{8}=GM.
\end{equation}
The solution provides an extremely simplified description; the radius contracts to exactly zero at the {\it collapse time} $\tcol=2\tta$.

Throughout the evolution, the density remains uniform, and the overdensity $\Delta=\rho / \bar{\rho}$ monotonically increases with time, and is parametrically described as follows:
\begin{equation}
\Delta=\frac{9}{2}\frac{(\theta-\sin\theta)^2}{(1-\cos\theta)^3}.
\label{deltaoftheta}
\end{equation}
Since $R$ contracts to zero at $t=\tcol$, $\Delta$ diverges to infinity,which is, of course, unrealistic. Instead, the system is usually assumed to instantaneously reach to the virial equilibrium ($\tv=\tcol=2\tta$). The radius of the sphere in this case is derived as follows.

At $t=\tta$, the kinetic energy $\Kta=0$ since all the matter inside the sphere stop expansion (turn-around). The kinetic energy $\Kv$ inside the sphere is assumed to be half the absolute value of the potential energy $\Wv$.  Since the density in the sphere is uniform at all times, the potential energy $W$ within the sphere should be given as
\begin{equation}
W=-\frac{3}{5}\frac{GM^2}{R}.
\end{equation}
Hence, by equating the total energies, one obtains
\begin{equation}
\Rv=\frac{1}{2}\Rta.
\label{RvtoRta}
\end{equation}
Note that the factor 3/5 in the potential energy depends in general on the density profile. It is essential that the retained uniformity of density yields the {\it same} factor of 3/5 both at $\tta$ and $\tcol$. We will revisit this point in Section 5.

The radius of the sphere {\it instantaneously} takes the finite value, $\Rv$, rather than vanishing, which is still a simplified assumption. Also, the sphere is often assumed to be totally isolated, so the size stays unchanged $\Rv$. In reality, however, the sphere grows due to the infall of surrounding matter.

The above results indicate that the overdensity $\Delta$ at $\tv$ is
\begin{equation}
\Delta(\tv)=18\pi^2\approx177.7
\end{equation}
This value is widely used as the threshold of halo identification in numerical simulations. Note that, this value is independent of the mass $M$ and the initial overdensity $\deltai$.

The above results are strictly correct only in the EdS universe, and we need to take into account the cosmological constant $\Lambda$. Even in the universe with $\Lambda$, the relation $\tcol=2\tta$ still holds, but both times are later than those of the sphere of the same $\deltai$ in the EdS universe, due to the existence of $\Lambda$.

In addition, the ratio $\Rv / \Rta$ becomes slightly less than 1/2 because the cosmological constant contributes to the total energy in the form of $-\Lambda MR^2/10$ (the factor 1/10 comes from the uniformity of density). The difference from 1/2 depends on the mean matter density parameter $\Omega_m$ at $t=\tv$. Along with the change of $\Rv / \Rta$, the virial overdensity $\Delta(t=\tv)$ (with respect to the cosmic mean density, not to the critical density) also depends on $\Omega_m$ at $t=\tv$. According to \cite{Lacey93}, \cite{Eke96} and \cite{Nakamura97}, $\Rv / \Rta\approx 0.483$ and $\Delta(t=\tv)=355.4$ for our simulated universe with $\Omega_{m,0}=0.279$, assuming that the sphere collapses exactly at $z=0$. We use the symbol $\Deltav$ to denote the value of 355.4 throughout the rest of this paper, and use the value in identifying halos from the N-body simulation (Section 3.2). 

In summary, TSC is based on several simplified assumptions including exact sphericity and uniformity, no velocity dispersion and instantaneous relaxation. We examine the validity of this highly idealized model by comparing against the evolution of {\it individual} halos in the following sections.

\section{Halos from cosmological N-body simulation}

\subsection{Numerical Simulation and Halo Catalog}

We run a cosmological N-body simulation from $z=99$ to $z=0$. We use TreePM code {\tt Gadget-2} \citep{Springel05} and the initial condition is generated by {\tt MUSIC} code \citep{Hahn11}, which employs second order Lagrangian perturbation theory. The transfer function at the initial redshift $\zini=99$ is generated by the linear Boltzmann code {\tt CAMB} \citep{Lewis00}. We assume the flat $\Lambda$CDM model with the cosmological parameters consistent with the {\it Wilkinson Microwave Anisotropy probe} (WMAP) 9 year result \citep{Hinshaw13}: ($\Omega_{\rm m,0}$, $\Omega_{\Lambda, 0}$, $h$, $n_s$, $\sigma_8$)=(0.279, 0.721, 0.7, 0.972, 0.821). We use a periodic box of $360\ h^{-1}\mathrm{Mpc}$ on a side in comoving scale. The number of dark matter particles is $1024^3$ and the mass of each particle is $m_p = 3.4 \times 10^{9} h^{-1} M_\odot$. The gravitational softening length is fixed at $20\ h^{-1} \mathrm{kpc}$ comoving.

To identify halos in our simulation, we follow the friends-of-friends (FOF) algorithm \citep{Davis85}. We set the linking parameter $b=0.159$, corresponding to the virial density $\Deltav=355.4$. Some authors \citep{More11, Courtin11} have pointed out that the overdensity within a FOF halo varies with its mass even for the same linking parameter $b$. In our study, however, this issue would not be important since the mass range of the simulated halos is relatively narrow and a ``halo'' is redefined using the spherical overdensity. 

We obtain 17535 halos with mass $> 10^{13} h^{-1} M_\odot$ at $z=0$, and we choose to analyze 100 most massive halos with mass range 2.06 $<M / (10^{14}h^{-1}M_\odot)<$ 16.6 in order to have a good mass resolution for each halo. In particular, Figures \ref{dist} - \ref{dists} below utilizes six halos listed in Table \ref{sixhalos} for illustration.

\begin{table}[h]
\begin{center}
\begin{tabular}{lccc}
\hline
Name & $M$ & $\deltai$ & $z_{\rm ta}$\\
& [$10^{14} h^{-1} M_\odot$] & [$10^{-2}$] &\\
\hline
Halo I & 16.6 & 2.72 & 1.12\\
Halo II & 16.3 & 2.61 & 1.21\\
Halo III & 8.38 & 2.86 & 0.870\\
Halo IV & 4.87 & 2.36 & 0.891\\
Halo V & 4.13 & 2.79 & 1.04\\
Halo VI & 2.65 & 3.06 & 1.28\\
\hline
\end{tabular}
\end{center}
\caption{\textup{Six halos selected for Figures \ref{dist} - \ref{dists} below. The turn-around redshift is estimated from the maximum point of the cubic-spline interpolated radius $R(z)$ of the simulated halo from the fifteen redshift data (see text in Section 3.2).}
\label{sixhalos}}
\end{table}

\subsection{How to Trace Back the Evolution of Simulated Halos}

Using the above simulation data, we compare the evolution of each halo with TSC. To this end, we define a sphere that corresponds to a halo identified by the FOF algorithm at $z=0$ as follows. First we calculate the center-of-mass of the FOF member particles of a halo. Starting from the center-of-mass, we find the radius within which the overdensity becomes $\Deltav=355.4$. Then we calculate the center-of-mass of all the particles in the sphere (not necessarily the FOF member particles), and repeat the procedure again. This process is iterated until the center-of-mass position of the particles in the sphere matches to the center of the sphere within 1 $h^{-1}$ kpc in comoving coordinates. After the iteration converges, we obtain the mass and radius of that halo, $M_{355}$ and $R_M(z=0)$, respectively.

For the purpose of our current study, we need the protohalos at different redshifts, which correspond to the halos identified at $z=0$ according to the above procedure. We have stored fifteen simulation snapshots; $z_i=99$ (initial), 49, 9, 5, 4, 3, 2, 1.5, 1, 0.8, 0.6, 0.4, 0.2, 0.1 and 0. We use them to trace back the evolution of halos identified at $z=0$.  For each redshift $z=z_i (\neq0)$, we trace the distribution of the FOF member particles back to $z_i$, and calculate the center-of-mass of the distribution at $z_i$ . From the center-of-mass, we find the radius inside which the mass is equal to $M_{355}$ computed from each halo at $z=0$. The iteration process described above is carried out to determine the center of the sphere. Finally, we define $R_M(z_i)$ as the radius of the sphere. Note that the mass $M_{355}$ is constant while $R_M(z_i)$ is a function of redshift. This procedure mimics the description of TSC.


The left panels of Figure \ref{dist} plot all the particles within the thickness of 0.03 $R_M(z)$ at each redshift around its center of mass. In order to clarify the degree of mixing of the particles, they are plotted in different colors according to their initial positions (at $z=99$). By $z=9$, the particles are little mixed, and the particles with the same color roughly keep the shape of the spherical shell. By $z=3$, however, filament-like structures already started to emerge, indicating that the uniform density assumption begins to break.

The colors of the particles at z=1, 0.4 and 0 are defined according to their positions at $z=1$. The redshift $z=1$ approximately corresponds to the turn-around epoch of Halo I. The colors are violently mixed by $z=0.4$ (after about 3.5 Gyr), in contrast to the period from the initial time to $z=2$ (about 3.3 Gyr). This is because the relaxation has finished in the inner part of the halo, and particles in the relaxed region are stirred due to shell-crossing.

\subsection{Evolution of Velocity Dispersion in Phase Space}

In order to see the degree of particle-mixing more clearly, we consider the phase space as well. Note that, throughout this paper, we refer to the space of radial coordinate and radial velocity ($r$, $v_r$) as the ``phase space'', for the comparison with the spherical model \citep[e.g., ][]{Colombi15,Sousbie15}. The right panels of Figure \ref{dist} demonstrate the phase space distribution of particles colored in the same way as the left panels. The right panels plot a randomly selected one percent of the particles inside the sphere within the shown radial range while the left panels plot all the particles in the slice region.

By definition, the initial particle distribution looks like seven color bars. These bars become gradually tilted by $z\sim2$. Although the coordinate space distribution has exhibited clear non-sphericity by $z=3$, the phase space distribution looks still well ordered. The innermost particles, however, have fallen into the center and then have positive (outward) radial velocity. By $z=2$, the region with large velocity dispersion $\sigma_r^2$ has formed. The region gradually expands outward, and finally reaches outside the halo radius $R(z=0)$. This is the most remarkable difference between the simulation and the description of TSC.

To look into the region with large $\sigma_r^2$ more carefully, we fully exploit all the particles and visualize the phase space density in Figure \ref{colcont}. The figure clearly shows the high density region around the outer end of the large $\sigma_r^2$ region, indicating the stream of the particles that have (more than) once fallen into the center. Such a motion of particles creates the large $\sigma_r^2$ region expanding outward.

It is informative to consider here the prediction of the self-similar model for the EdS universe \citep{Filmore84, Bertschinger85}. In the self-similar model, a spherical shell falls toward the center, and moves outward again after shell-crossing. The shell turns-around at some radius and falls back toward the center. Such oscillations of a number of  shells account for the development of the velocity dispersion in a halo. The physical size of the halo increases with time as more shells infall with larger turn-around radius. 

The above picture explains, at least qualitatively, the evolution of $\sigma_r^2$ of Halo I shown in Figure \ref{colcont}. From $z=1$ to $z=0$, the velocity dispersion develops from the center of the halo. The profile of $\sigma_r^2$ shows a sharp drop-off at the radius corresponding to the end of the large velocity dispersion region. In contrast, the radial (peculiar) velocity $v_r$ almost vanishes in the central region, while it is negative in the outer region, representing the falling particles. All these features are consistent with the self-similar model.

In addition to the ``regular'' development of $\sigma_r^2$ described above, the inhomogeneity contributes to the evolution of $\sigma_r^2$ in the simulated halos. For example, the infall and the subsequent turn-around of a substructure generates an additional velocity dispersion that is not described in the self-similar model. Such a process enhances individuality of halos, and makes it difficult to find universality (if any) of the evolution of $\sigma_r^2$, as will be discussed again in later sections.

For the phase space distribution at $z=0$, the solution of the self-similar model is overplotted. The overall feature of the simulated halo is followed, at least qualitatively, by the self-similar solution. Especially, groups of the particles in course of the first and second turning-arounds are apparent in the phase space
distribution. Strictly speaking, however, the self similar model describes a {\it spherical} halo in the {\it EdS universe}, which naturally leads to the big difference between the simulation data and the overplotted self-similar solution. In addition, the density profile of the self-similar solution is predicted to be asymptotically proportional to $ r^{-9/4}$, which is inconsistent with the NFW density profile \citep{Navarro95, Navarro96, Navarro97}. Therefore the self-similar solution is not fully reliable when we {\it quantitatively} investigate the evolution of velocity dispersion.

\cite{Adhikari14} have also considered the region with large velocity dispersion in a different context. They refer to the radius where the density sharply drops as ``splashback radius'', which is essentially the same as the locus where the velocity dispersion sharply drops. They proposed the splashback radius as a more physically motivated definition of a dark halo, instead of the traditional definition by using some threshold overdensity. Although the splashback radius is often much beyond the X-ray observed region of galaxy clusters, \cite{More15} have shown that the relation between the splashback and $R_{200 \rm m}$, within which the overdensity is 200 times the mean matter density, can be written as a function of the peak height inside $R_{200 \rm m}$. Furthermore, \cite{More15} indicated that the splashback radius may be already observed as a caustic of line-of sight velocity of galaxies by \cite{Rines13}. Although \cite{Adhikari14} claimed that it is difficult to unambiguously determine the splashback radius of individual halos, these studies indicate the importance of velocity dispersion in the halo evolution.

\section{Comparison of Halo Radius Evolution Against TSC and Spherically Averaged Jeans Equation}

We now compare the evolution of the sphere characterized by the radius $R_M(z)$ defined in Section 3.2, with the prediction of TSC. Figure \ref{collapse} demonstrates the results for the six halos in Table \ref{sixhalos}. The TSC predictions (black solid line) are calculated from the initial overdensity of each halo. From the initial time until shortly before the turn-around epoch $\zta$, $R_M(z)$ is very close to the model prediction, despite the fact that non-sphericity and non-uniformity develop by $z\sim3$. From around $\zta$, $R_M(z)$ deviates from the model prediction; the turn-around epoch is delayed, and thereafter the radius of the simulated halo becomes systematically larger than the model. Finally, the radius $R_M(z)$ does not collapse to zero (naturally), but settles into a finite radius. In addition, the present radius is also larger than the model prediction. Although the degree of the deviation from TSC varies from halo to halo, the above trend holds for majority of the simulated halos: the simulated halos turn around later, and have larger radii both at $\zta$ and $z=0$ than those predicted by TSC.


We suspect that the difference between the simulation and TSC is mostly due to the velocity dispersion focused on in the previous section. In numerical simulations, the motion of dark matter particles should be described not by Equation (\ref{eomtsc}), but by the (three-dimensional) Jeans equation. We here focus only on their radial motion to see the effect of the velocity dispersion in the framework of spherical symmetry. The spherically symmetric version of the Jeans equation is
\begin{equation}
\frac{Dv_r}{Dt}
=-\frac{1}{\rho}\frac{\partial (\rho\sigma_r^2)}{\partial r}
-\frac{2\sigma_r^2-\sigma^2_{\rm tan}}{r}-\frac{GM}{r^2},
\label{Jeans}
\end{equation}
where $D/Dt$ denotes the Lagrangian differentiation, and $\sigma^2_{\rm tan}$ is the tangential velocity dispersion of dark matter. Note that $\sigma^2_{\rm tan}$ includes dispersions in two directions ($\theta$ and $\varphi$ directions in the spherical coordinates). The density and velocity dispersion usually decrease as a function of radius, so the first term is expected to delay the collapse epoch. To confirm this, we evaluate the first two terms at $r=R_M(z)$ in the right-hand-side of Equation (\ref{Jeans}) from the simulation data for each of the fifteen redshifts, and solve the Equation (\ref{Jeans}) with the two terms replaced by using the cubic-spline interpolated values from the fifteen redshifts.

The results are illustrated by the dashed lines in Figure \ref{collapse}. The prediction based on Equation (\ref{Jeans})reproduces the simulation results much better than that of TSC at least for Halos I, II, IV and V. Therefore, we confirm that velocity dispersion explains the delayed turn-around and the stopped contraction. For Halos III and VI, on the other hand, the modification is not so successful. This is probably attributed to the strong non-sphericity of the dark matter distribution. In fact, as shown in Figure \ref{dists}, Halo III has undergone a drastic merger of two similar mass objects. Halo VI, on the other hand, does not undergo such a big merger, but its mass accretion occurs prominently along a single direction. In the other four halos, matter assembles around the central structure from every direction (The distribution of Halo I is shown in Figure \ref{dist}).

Hence, the level of the improvement depends on the sphericity in the evolution of the halos. Also, the velocity dispersion terms in Equation (\ref{Jeans}) are not uniquely determined, so the details of the result depend on their evaluated values. Our present purpose, however, is not to precisely improve TSC, but to confirm the effect of velocity dispersion. Thus we conclude that the velocity dispersion plays an important role in the halo evolution from the above comparison.

If we can model the evolution of $\sigma_r^2$ of an individual halo fully from initial conditions, such a model helps us understand the halo evolution beyond the spherical collapse model. We have found, however, that the profiles of $\sigma_r^2$ and density vary sensitively from halo to halo: Although we made sure that $\sigma_r^2$ calculated from the linear power spectrum of the matter density fluctuations can approximate that of simulated halos at the early stage ($z\gtrsim5$), the late evolution strongly depends on merger and mass accretion processes of each halo. Also, as stated in Section 3.3, the evolution of $\sigma_r^2$ in the self-similar model is not quantitatively successful in describing that of the simulated halo. Thus the improved modelling, we do not attempt here, remains as an important future work.

As a preliminary step to the above challenge, therefore, we evaluate more quantitatively the deviation from TSC. In the next section, we do so by comparing the radii of the halos at the initial, the turn-around, and the present times.

\section{Effect of Velocity Dispersion on Prediction for Present Radius}

In this section, we focus on the difference in the present radius $R_0$ between the simulation and TSC prediction. We first compare $R_0$ and $\Rta$ of the simulated halos. Next, we compare $R_0$ of the simulated halos with the predicted value by TSC from the initial conditions. To investigate the origin of the difference, it is essential to factorize the ratio into energy terms, and we here provide relevant definitions for preparation.

The ratio of radii at two epochs may be predicted from the conservation of the total energies $E$ at those epochs. We decompose the total energy $E$ into the kinetic energy $K$ (due to both the Hubble expansion and the peculiar velocity), the gravitational potential energy $W$ and the energy $W^\Lambda$ due to the cosmological constant.

The potential energies $W$ and $W^\Lambda$ within the sphere of mass $M$ and radius $R$ are given by
\begin{equation}
W=-\gamma\frac{GM^2}{R}
\end{equation}
and
\begin{equation}
W^\Lambda=-\gamma^\Lambda\Lambda MR^2,
\end{equation}
respectively, where the parameters $\gamma$ and $\gamma^\Lambda$ depend on the density profile inside the halo (see, e.g., Equation \ref{eqgamnfw} and discussion therein). Even if the particle distribution is non-spherical, the above parametrization is valid as long as a sphere of radius $R$ is considered.

In order to derive a radial ratio at two epochs, the kinetic energy $K$ must be associated with the other energy terms. In the settings of TSC, $K=0$ at $z=\zta$, and $K=-(1-5\deltai /3)W$ at $z=\zini$, In order to consider the difference from these predictions, we define the parameter $\alpha$ as
\begin{equation}
\alpha=-\frac{K}{W}.
\end{equation}
Then the total energy $E$ can be written as
\begin{eqnarray}
E &=& K+W+W^\Lambda \nonumber\\
&=& -(1-\alpha)\gamma\frac{GM^2}{R}-\gamma^\Lambda \Lambda MR^2, 
\end{eqnarray}

From now on, in order to represent the above quantities at different epochs, we use the subscript $X$ to mean either of ``ini'', ``ta'', and ``0'', denoting the quantity at $z=\zini$, $\zta$ and 0 (present), respectively.

In the following sections, we use the ratio between the radii at $z=0$ and another epoch, assuming that the virial theorem is applicable to the halos at $z=0$.  By equating the total energies $E_X$ ($X=$ ``ta'' or ``ini'') and $E_0$, one obtains the ratio $R_0/R_X$ in terms of those coefficients of the energies. Strictly speaking, $R_0/R_X$ is a solution of the cubic equation, and the exact expression is not useful in understanding how each energy term contributes the difference between the simulation and TSC. The contribution of $\Lambda$ is, however, very small compared to $W$:
\begin{equation}
\frac{W^\Lambda}{W} =\frac{6}{\Delta}
\frac{\gamma^\Lambda}{\gamma}\frac{\Omega_\Lambda}{\Omega_m},
\end{equation}
which is, for example, less than 1 \% at $z=0$ ($\Delta=355.4$ by definition for our adopted cosmology). Hence we can treat $W^\Lambda/W$ as an infinitesimal.

Note, however, that, according to the virial theorem in the universe with $\Lambda$ \citep{Nowakowski02}, a virialized halo satisfies $-K/W=1/2 - W^\Lambda/W=0$ in the same settings as TSC, i.e., $\alpha'_0$ includes an additional $W^\Lambda/W$. Hence we define another parameter $\alpha'_0$ only for $z=0$ as
\begin{equation}
\alpha'_0=-\frac{K_0}{W_0}+\frac{W^\Lambda_0}{W_0},
\end{equation}
although the difference between $\alpha_0$ and $\alpha'_0$ is negligible at the level of the following discussion.

Then the total energy $E_0$ at $z=0$ is $E_0=$ $(1-\alpha'_0)W+2W^\Lambda$.  By equating $E_0$ and $E_X=$ $(1-\alpha_X)W_X+W^\Lambda_X$, we solve $R_0/R_X$ perturbatively up to the leading term in $W^\Lambda/W$ ($\propto \Lambda	R^3/(GM)$). The result is
\begin{equation}
\frac{R_0}{R_X}=\frac{1-\alpha'_0}{1-\alpha_X}
\frac{\gamma_0}{\gamma_X}\frac{1}{\beta_X^0}(1-\epsilon^0_X),
\label{RYtoRX}
\end{equation}
where
\begin{equation}
\beta_X^0=\frac{E_0}{E_X}
\end{equation}
and	
\begin{eqnarray}
\epsilon^0_X&=&\left[\frac{\gamma^\Lambda_X}
{\gamma_X(1-\alpha_X)}
-\frac{2\gamma^\Lambda_0}{\gamma_0(1-\alpha'_0)}
\left(\frac{\gamma_0(1-\alpha'_0)}{\gamma_X(1-\alpha_X)\beta_X^0}\right)^3
\right]\nonumber\\
&&\times \frac{\Lambda R_X^3}{GM}.
\end{eqnarray}
A substantial fraction of the particles in a simulated halo defined in Section 3.2 indeed move into and out of the sphere between the two epochs, and the total energy within the sphere is not necessarily guaranteed to be conserved. The parameter $\beta$ indicates the degree of the energy conservation.

We calculate the above parameters $\alpha$, $\beta$ and $\gamma$ for the 100 simulated halos. For a simulated halo, the kinetic energy $K$ is calculated as
\begin{equation}
K=\frac{1}{2}\sum_im_i(\bm{v}_i+H\bm{x}_i)^2,
\end{equation}
where $m_i$, $\bm{x}_i$ and $\bm{v}_i$ are mass, position and peculiar velocity of the $i$-th particle, and $H$ is the Hubble parameter at the epoch. The summation is taken over all the particles within the sphere of radius $R$.

The gravitational potential energy $W$ is calculated as
\begin{equation}
W=-G\sum_{i<j}\frac{m_im_j}{|\bm{x}_i-\bm{x}_j|},
\end{equation}
where the summation is taken over all the combinations of the $i$-th and $j$-th particles within the sphere of radius $R$. The parameter $\gamma$ is simply computed as $W/(GM^2/R)$.

\subsection{Comparison of $R_0$ and $\Rta$}

The TSC prediction (\ref{RvtoRta}) is based on the energy conservation between the turn-around and the collapse epochs. As seen in Section 4, however, it is difficult to define unambiguously its collapse time. So, we first compare the {\it present} radius $R_0$ with the turn-around radius $\Rta$ instead. If TSC is exact, $R_0$ defined with $\Deltav=355.4$ should correspond to $\Rv$ for objects that collapsed at $z=0$. Note that, both $R_0$ and $\Rta$ in this section are of the simulated halos.

The ratio $R_0/\Rta$ predicted by TSC is 0.483, which can be compared with the simulation. The top-left panel of Figure \ref{tato0} shows that the $R_0/\Rta$ of the simulated halos is 0.56 on average, with roughly 10 - 20 percent scatter. This level of the deviation may be fully expected, given the extremely simplified assumptions of TSC.

In order to identify the origin of the discrepancy more quantitatively, we use Equation (\ref{RYtoRX}). In TSC, $\epsilon^0_{\rm ta}=0.032$, which explains the difference between the values of $R_0/\Rta$ between the EdS universe ($R_0/\Rta=0.5$) and the universe with $\Lambda$ ($R_0/\Rta=0.483$). For simplicity, we do not consider the contribution of each parameter in $\epsilon^Y_X$ to $R_0/\Rta$, and use the following:
\begin{equation}
\frac{R_0}{\Rta}=0.483\frac{1-\alpha'_0}{0.5}
\frac{1}{1-\alphata}\frac{\gamma_0}{\gammata}\frac{1}{\betata}.
\label{R0toRta}
\end{equation}
Note that the contribution of $\Lambda$ is partly incorporated in $\betata$. TSC predicts that the kinetic energy vanishes at $\zta$ ($\alphata=0$), and the virial theorem states that $\alpha'_0=1/2$. In addition, the density profile is always uniform ($\gammata=\gamma_0=3/5$). Thus, combined with energy conservation ($\betata=1$), one obtains $R_0/\Rta=0.483$.

The number of our available snapshots of the simulation is limited, so we define the energy terms of each halo at $\zta$ as follows. First, for a simulated halo, the values of radius at fifteen redshifts are cubic-spline interpolated, and its maximum value and the corresponding epoch are defined as $\Rta$ and $\tta$, respectively. We calculate the energy terms for the two snapshots bracketing $\tta$, and define the energy at $\tta$ with the linear-interpolation.

Let us consider $\betata$ first. The top-right panel of Figure \ref{tato0} shows the calculated $\betata$ for 100 halos. The average $\langle\betata\rangle$ is 0.96, so the total energy is conserved to a good approximation. This is not trivial, since a significant fraction ($\sim 20$ \%) of the particles in the sphere is changed. While we take into account the factor, $\betata$ does not play a major role in Equation (\ref{R0toRta}).

The middle panels in Figure \ref{tato0} plot $\alphata$ and $\alpha'_0$. While TSC states that $\alphata=0$, the simulated halos have roughly $\alphata=0.37$ on average. This is the largest deviation from TSC, and can be attributed to the velocity dispersion in the central region of halos; at the turn-around, the halo expansion velocity of the outer shell almost vanishes, but the velocity dispersion of the inner region significantly contributes to the kinetic energy.

At the present time, $\alpha'_0$ is 0.62 on average, which is larger than 1/2 predicted by the virial theorem. This result can be understood according to \cite{Lokas01}, who study the virialized state of a spherical halo with the NFW density profile. They defined the virialized state as a solution of the spherical Jeans equation, and derived the ratio $K/W$ as a function of radius with the concentration parameter, $c$, and parametrized velocity-anisotropy. Since this model is for the equilibrium state, the halo has no average velocity, but finite velocity dispersion, which yields the substantial kinetic energy. For any concentration parameter and velocity-anisotropy, they find that $K/W$ is larger than 1/2 at any radius, and increases toward the center. This is mainly due to the density and velocity dispersion inside the sphere, and the matter surrounding the halo is not important. In most cases, $K/W$ at the virial radius is in the range from 0.5 to 1, which is in qualitative agreement with our simulated halos. The difference in $\alpha'_0$ between the simulation and TSC is also attributed to the velocity dispersion that is naturally expected in the inside-out collapse model in the CDM universe.

Next, we look at $\gammata$ and $\gamma_0$, which are shown in the bottom panels in Figure \ref{tato0}. Both are distributed around unity, which is different from 3/5 for the uniform density profile. As stated before, $\gamma$ depends on the density profile inside the sphere. For example, the single power-law density profile $\rho\propto r^{-p}$ ($p<5/2$) results in $\gamma=(3-p)/(5-2p)$. Hence $\gamma=1$ implies $p=2$.

For the NFW profile with the concentration parameter $c$, we obtain
\begin{equation}
\gamma=c\left[\frac{c(2+c)}{2(1+c)^2}
-\frac{\log(1+c)}{1+c}\right]\left[\log(1+c)-\frac{c}{1+c}\right]^{-2}
\label{eqgamnfw}
\end{equation}
at the virial radius. Figure \ref{gammaNFW} plots Equation (\ref{eqgamnfw}), showing that $\gamma$ is a increasing function of $c$ and $\gamma (c=0)=2/3$. Hence, for any $c$, the NFW profile predicts larger values of $\gamma$ than 3/5 from the uniform profile. According to \cite{Oguri12}, halos with the mass range $2. <M / (10^{14} h^{-1} M_\odot) <20$ (our sample) typically have $3<c<10$, implying $0.9<\gamma<1.3$. The range agrees well with our $\gamma_0$. At $\zta$, the density profile of the halo is not necessarily described by the NFW profile, so the above discussion can not be applied. The difference between $\gammata$ and $\gamma_0$ is small, so they do not play a major role. Also, as long as $R_0/\Rta$ is concerned, the deviation of $\gamma$ from 3/5 itself is not important, but difference at the two epochs contributes the budget.

In summary, the ratio $R_0/\Rta=0.48$ is increased by $(1-\alphata)^{-1}=1.6$ (not 1), and decreased by $(1-\alpha'_0)/0.5=0.76$ relative to the TSC prediction, which finally yields 0.58, approximately explaining the mean value of 0.56. This implies that, although the non-zero velocity dispersion effect is fairly large, the other effects tend to cancel it in practice.

\subsection{Comparison of $R_0$ and $\Rini$}

An important advantage of TSC is its definite prediction for evolution of a halo {\it from the initial condition} and also the insensitivity of the result to $M$. Hence, we now compare the TSC prediction of the halo radius $\Rmod$ at $z=0$ from the initial condition ($z=99$) measured for the simulation, against the radius $\Rsim$ measured for simulated halos $z=0$. From now, we distinguish the two by denoting ``TSC'' or ``sim''.

The upper-left panel of Figure \ref{inito0} plots the ratio $\Rsim/\Rmod$ for the 100 simulated halos. For most  halos, $\Rsim/\Rmod$ is greater than unity. We suspect that the velocity dispersion produces this trend, and investigate its effect in the following.

We again use the spherical collapse model to describe $\Rmod$ for simplicity. In the linear regime ($\theta \ll 1$), Equations (\ref{Roftheta}) and (\ref{deltaoftheta}) are approximated as $R\approx \Rta\theta^2/4$ and $\delta\approx3\theta^2/20$. Combined with $R_0=\Rta/2$, the model prediction for the present radius is given by
\begin{equation}
\Rmod=\frac{3}{10}\deltai^{-1}\Rini.
\end{equation}
(Here we have derived the above expression based on TSC in the EdS universe, but it holds in the flat universe up to the first order of $W^\Lambda/W$.)
Using the ratio $\Rsim/\Rini$ that can be written in the form of Equation (\ref{RYtoRX}); we obtain
\begin{eqnarray}
\frac{\Rsim}{\Rmod}&=&\frac{10}{3}\deltai\frac{\Rsim}{\Rini}\nonumber\\
&=&\frac{10}{3}\deltai\frac{1-\alpha'_0}{1-\alphai}\frac{\gamma_0}{\gammai}\frac{1}{\betai},
\label{R0s}
\end{eqnarray}
where $\alphai$, $\betai$ and $\gammai$ are defined at $z=\zini$ as we did at $z=\zta$ (Section 5.1). Since $\Rini$ is very small, we neglect the correction due to $\epsilon^0_{\rm ini}$.

We calculate $\alphai$, $\betai$ and $\gammai$ from the simulation data. We again begin with looking at the energy conservation. We find that the particles within a halo change by about 30 \% from $z=\zini$ to $z=0$. As a result, the upper-right panel of Figure \ref{inito0} shows that the total energy within the sphere changes within by $\sim$ 7 \% from $z=\zini$ to $z=0$

At the initial time, the density is almost uniform, so $\gammai=3/5$, and $\alphai=1-5\deltai/3$ for small $\delta$. Actually, the simulated halos have the values of $\gammai$ and $\alphai$ very close to the theoretical values, as shown in the lower panels of Figure \ref{inito0}. Thus Equation (\ref{R0s}) practically reduces to
\begin{equation}
\frac{\Rsim}{\Rmod}=\frac{1-\alpha'_0}{0.5}\frac{\gamma_0}{0.6},
\end{equation}
which indicates that the deviation of $\Rsim/\Rmod$ from unity is largely dictated by $\alpha'_0$ and $\gamma_0$. If $\alpha'_0=1/2$ and $\gamma_0=3/5$, $\Rsim/\Rmod=1$. In reality, however, $(1-\alpha'_0)/0.5=$ 0.76 reduces $\Rsim/\Rmod$, and $\gamma_0/0.6=1.6$ increases $\Rsim/\Rmod$ to 1.3, which approximates the average $\langle\Rsim/\Rmod\rangle=$ 1.2. Therefore, the deviation $\Rsim/\Rmod$ is mainly attributed to the non-uniformity of the {\it present} density profile and the {\it present} kinetic energy due to the velocity dispersion.

It is interesting to see how the above results depend on the initial overdensity $\deltai$ of the simulated halos since the TSC predictions are almost independent of the halo mass and mainly determined by $\deltai$.

We have defined the spherical region for each halo based on the overdensity $\Deltav$. The value of $\Deltav$ corresponds to a halo that is predicted to collapse exactly at $z = 0$ by TSC. Figure 3 shows, however, the simulated haloes collapse significantly earlier (although it is difficult to precisely determine when they collapse, since the radius does not shrink to zero). This implies that $\deltai$ of the simulated halos is larger than predicted by TSC.

In fact, the upper panel of Figure 8 shows that all the simulated halos have the initial overdensity (normalized by the linear growth factor) larger than the linearly extrapolated threshold of $\delta_{\rm c}$ = 1.67 for a halo that is expected to collapse at present in TSC. In addition, there seems a weak trend that mass is anti-correlated to $\deltai$; more massive halos have smaller $\deltai$. Although the statistical significance is not strong, this is consistent with the initial density distribution of random-Gaussian field first derived by Doroshkevich (1970).

Because of the above correlation of mass and $\deltai$, we attempt to replot $\Rsim/\Rmod$ and $R_0/\Rta$ now in terms of $\deltai$. Here we recall that we have defined the sphere for each present halo based on the common $\Deltav$. So, by definition, $\Rsim/\Rmod$ is  proportional to $((1+\deltai)/\Deltav)^{1/3}\times\deltai$ (cf. Equation (\ref{R0s})). In fact, the lower-left panel of Figure \ref{delini} shows that $\Rsim/\Rmod$ follows a single curve as expected.

Similarly to $\Rsim/\Rmod$, $R_0/\Rta$ should follow a single curve if TSC were an exact description of the evolution of simulated halos. In reality, however, their relation has a relatively larger scatter around the mean relation as shown in the lower-right panel of Figure \ref{delini}. This corresponds to the deviation from the TSC prediction at $z=\zta$. While the degree of the scatter may be related to the non-sphericity of each simulated halo, we were not able to identify a clear dependence of non-sphericity on $\deltai$. We also confirmed that there is no clear dependence on $\deltai$ in the parameters such as $\alphata$, $\gammata$, etc. A further study on non-sphericity of halos may need a precise non-spherical definition of the region of simulated halos, and a wider mass range of halos, which we plan to study and present elsewhere.

\section{Summary and Discussion}

We have quantitatively tested the limitations and reliability of the predictions of the top-hat spherical collapse model (TSC).  Our halo-to-halo comparison of the spherically-averaged dynamics of 100 halos identified from a cosmological N-body simualtion yields the following results.

\begin{enumerate}
\item
Even though the averaging and the dynamics do not commute, the overall predictions of TSC approximately describe the evolution of the simulated halos fairly well, in particular prior to their turn-around epochs.  In reality, however, the non-uniformity/inhomogeneity of dark matter density profiles and the non-zero velocity dispersions, both of which are neglected in TSC, turn out to play an important dynamical role.
\item
Unlike a simplified TSC picture of the instantaneous collapse, dense clumps inside a halo collapse first, and merge and fall into the central region.  Thus a large velocity dispersion is developed from the center to outer parts.  The region with the velocity dispersion expands outward, and finally reaches outside the ``virialized'' region predicted by TSC.
\item
The velocity dispersion inside the halos strongly affects the size of the present {\it virial} radius of halos.  At the turn-around epoch $\zta$, the kinetic energy $K$ amounts to 37 \% of the gravitational potential energy $W$, which increases the ratio $R_0/\Rta$ of the radii of the simulated halos at $z=0$ and $z=\zta$ by $\sim$ 58 \%. The velocity dispersion also contributes to the kinetic energy, and $K/W$ becomes 0.62 on average (larger than 0.5 in TSC), which decreases $R_0/\Rta$ by 25 \%. In total, $R_0/\Rta$ is 0.56 on average, which is larger than the TSC prediction ($\approx$ 0.483) by 16 \%.
\item
Moreover, the ratio $\Rsim/\Rmod$ of the present radius of the simulated halos to the TSC prediction significantly deviates from unity; $\Rsim/\Rmod$ is 1.2 on average. The deviation from the TSC prediction is explained on average by $\sim$ 20 \% decrease due to the non-zero velocity dispersion effect, and $\sim$ 60 \% increase due to the non-uniformity of dark matter density profile at $z=0$.  While the two effects tend to cancel each other, those two effects need to be properly taken into account in the dynamical description of evolution of actual individual halos in the CDM universe.
\end{enumerate}

The interpretation of the present results may slightly vary.  It could be concluded that despite the idealized assumptions, the overall TSC predictions are largely reliable within 20 \%. On the contrary, however, we could argue that the TSC predictions should be critically re-examined given their wide applications in the modern precision cosmology.

In the latter spirit, the velocity dispersion, which is usually neglected in the standard TSC, needs to be discussed and modelled. Especially, its characteristic evolution in the phase space may provide a more physically reasonable definition of dark halos \citep{Adhikari14}, instead of the conventional definition on the basis of the overdensity threshold motivated by TSC.

As we have shown, the size of the simulated halos differs from the TSC predictions by $\sim$ 20 \%, which suggests that the conventional value of the threshold overdensity $\Deltav=355.4$ should be revised accordingly.  For this value of $\Delta$, in TSC, halos are supposed to collapses exactly at $z=0$. As seen in Figure \ref{collapse}, however, the predicted collapse time (black solid line) is significantly earlier than the present time, meaning that the mapping of the overdensity from the initial time to the present time is different from the Press-Schechter description. While \cite{Sheth02} succeeded in modifying the halo abundance by incorporating the halo ellipticity, the above result may also lead some modification of the halo abundance in the framework of the spherical symmetry.

Our halo identification in the present paper still adopts this value, and further discussion of this modification is beyond the current analysis; the quantitative implications have to be considered using a sort of iterative approaches of the comparison with different sets of halos identified by varying the values of $\Deltav$, which is very expensive and time-consuming. We plan to perform the analysis along this line and to present the results elsewhere.

To carry the above discussion further, a model to describe the halo evolution more accurately than TSC will be helpful. In other words, the modelling of $\sigma_r^2$ is needed. This is a challenging task since the evolution of $\sigma_r^2$ is sensitive to the evolution of individual halos including merger processes.  Indeed the velocity dispersion effect may explain the possible discrepancy between the ellipsoidal collapse model and the simulations/observations as mentioned in Introduction; heavier halos are more spherical in theoretical models \citep{Bond96, Rossi11}, while simulations and observations have drawn the opposite conclusion \citep{Jing02,Despali14,Kawahara10,Oguri10}. The origin of this discrepancy is often attributed to the merger process. Since the evolution of $\sigma_r^2$ is strongly related to the merger process of individual halos, the further consideration of the evolution of $\sigma_r^2$ may elucidate the origin of the problem.

Our current analysis suggests that the velocity dispersion of dark halos may be a key to improve the simplified TSC picture.  This is not only a theoretically motivated problem, but would have a big impact on the cosmology with galaxy clusters in particular.  We hope that our result detailed here will act as a first step towards a model construction to improve the simple, but widely used, TSC predictions.

\section*{Acknowledgements}
We thank the an anonymous referee for many valuable comments. This work is supported by JSPS Grantin-Aid for Scientific Research No. 26-11473 (D.S.), No. 25400236 (T. K.) and No. 20340041 (Y. S.). K.O. is supported by Advanced Leading Graduate Course for Photon Science. Numerical simulations were carried out on Cray XC30 at the Center for Computational Astrophysics, National Astronomical Observatory of Japan.

\begin{figure*}[h]
\begin{center}
\includegraphics[width=6cm]{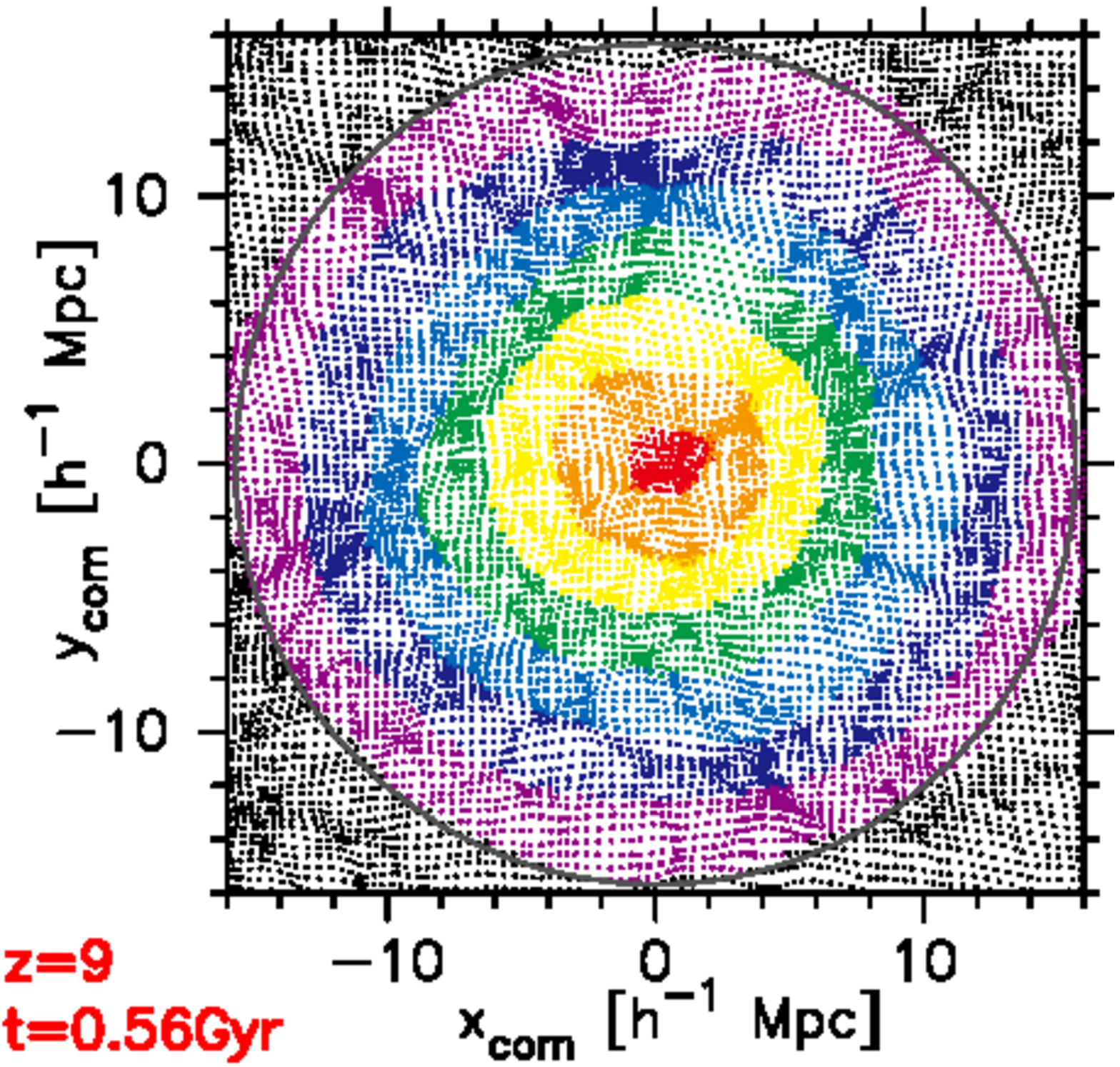}\qquad
\includegraphics[width=6cm]{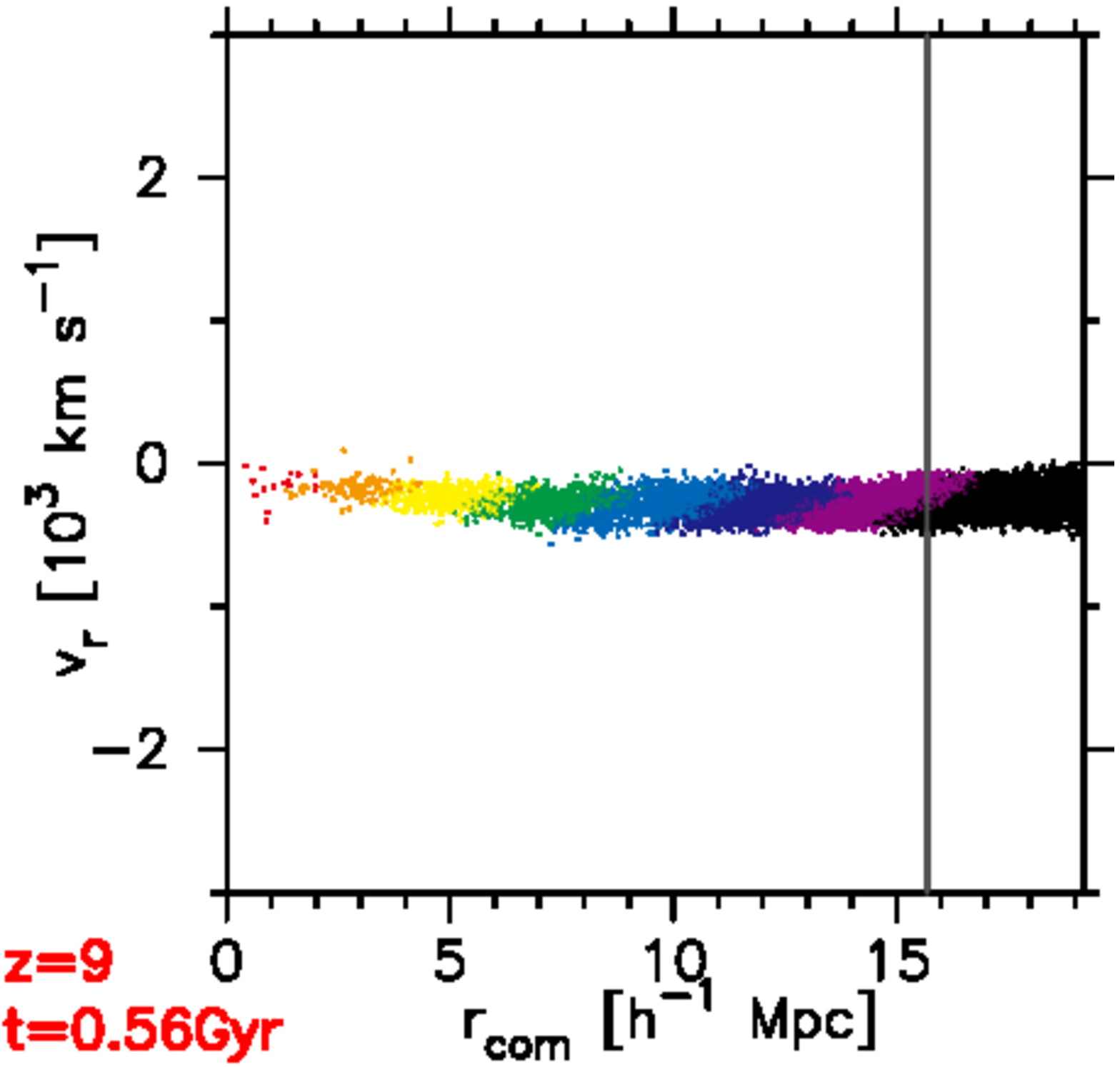}\\
\includegraphics[width=6cm]{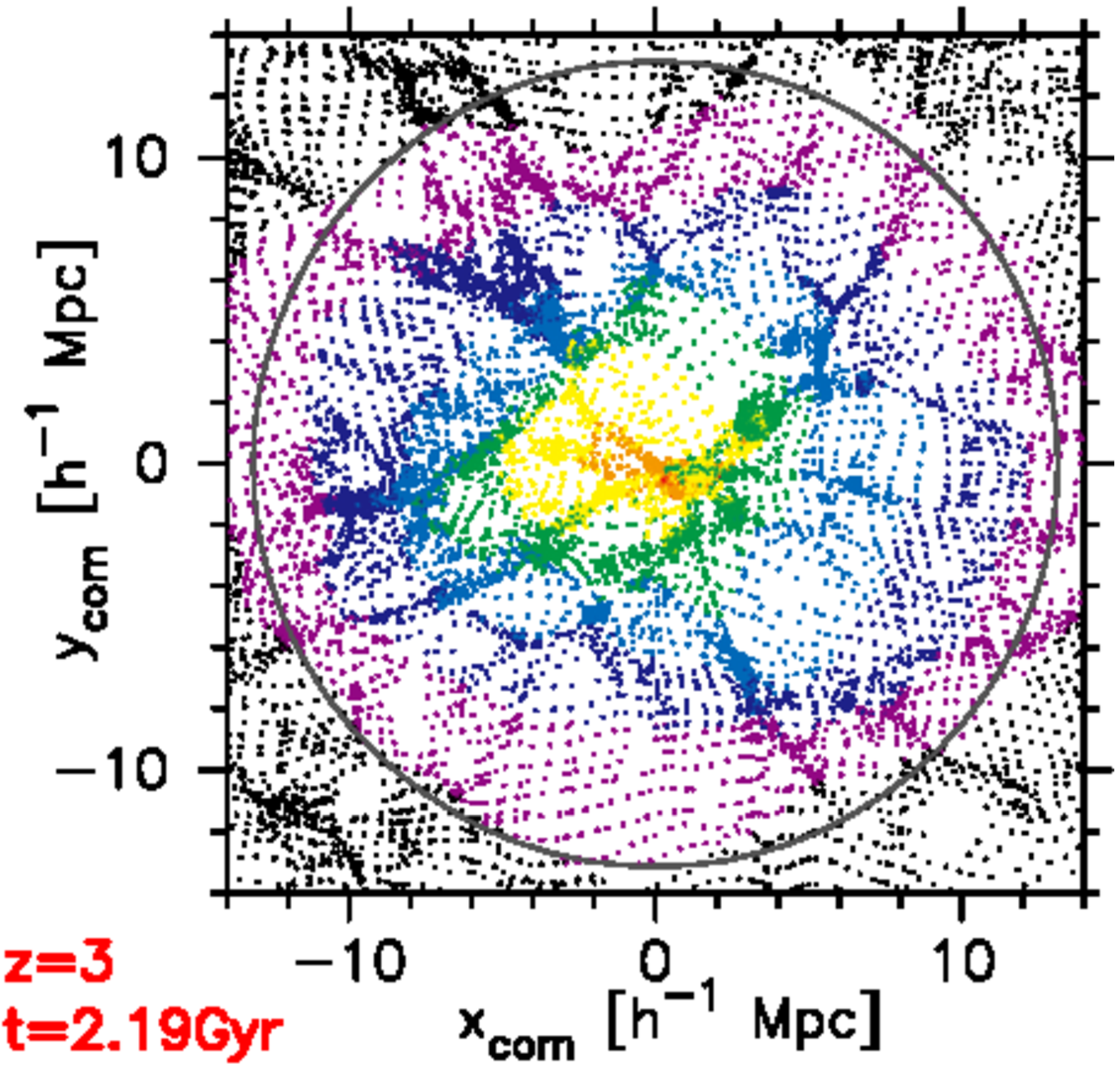}\qquad
\includegraphics[width=6cm]{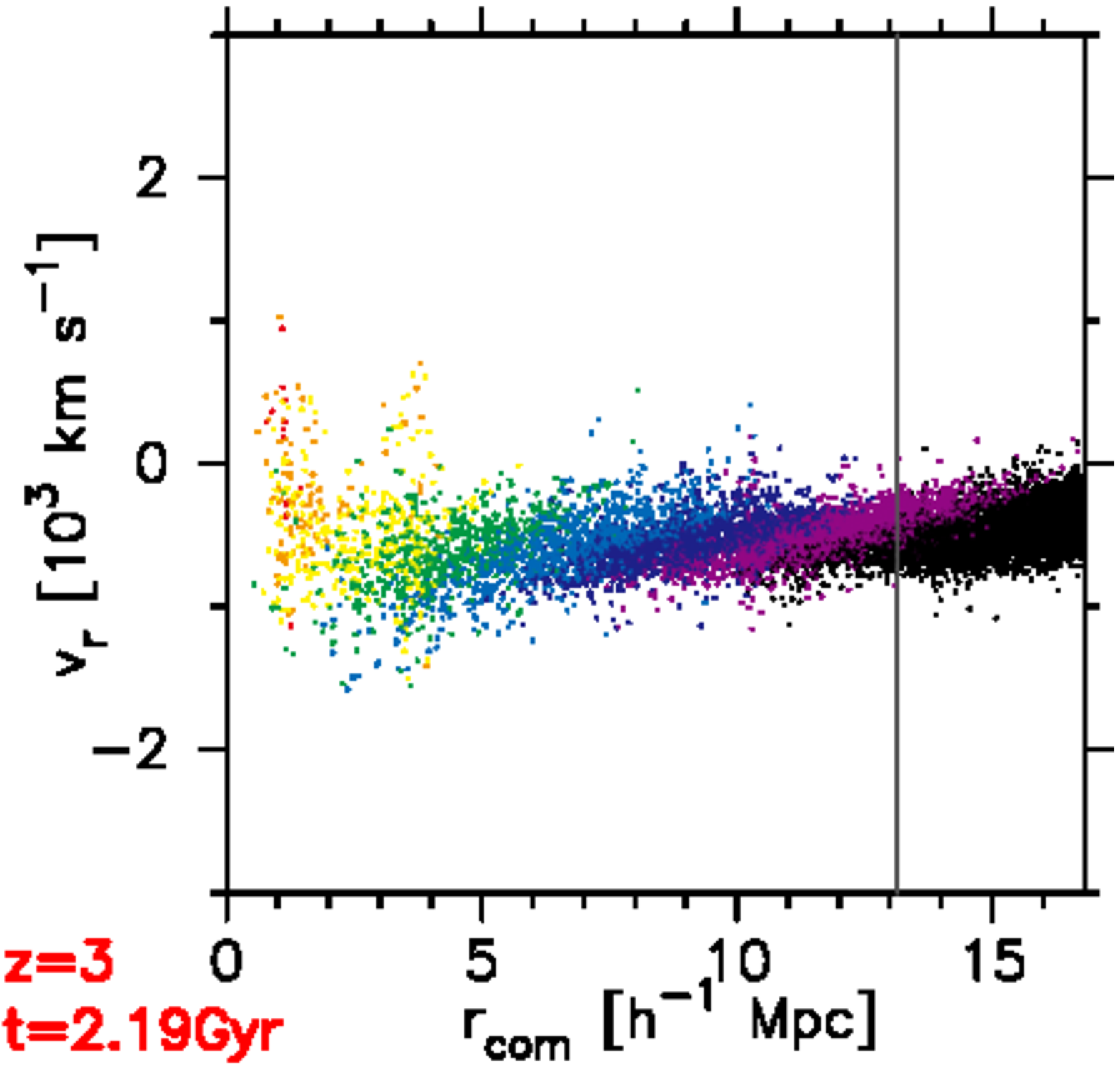}\\
\includegraphics[width=6cm]{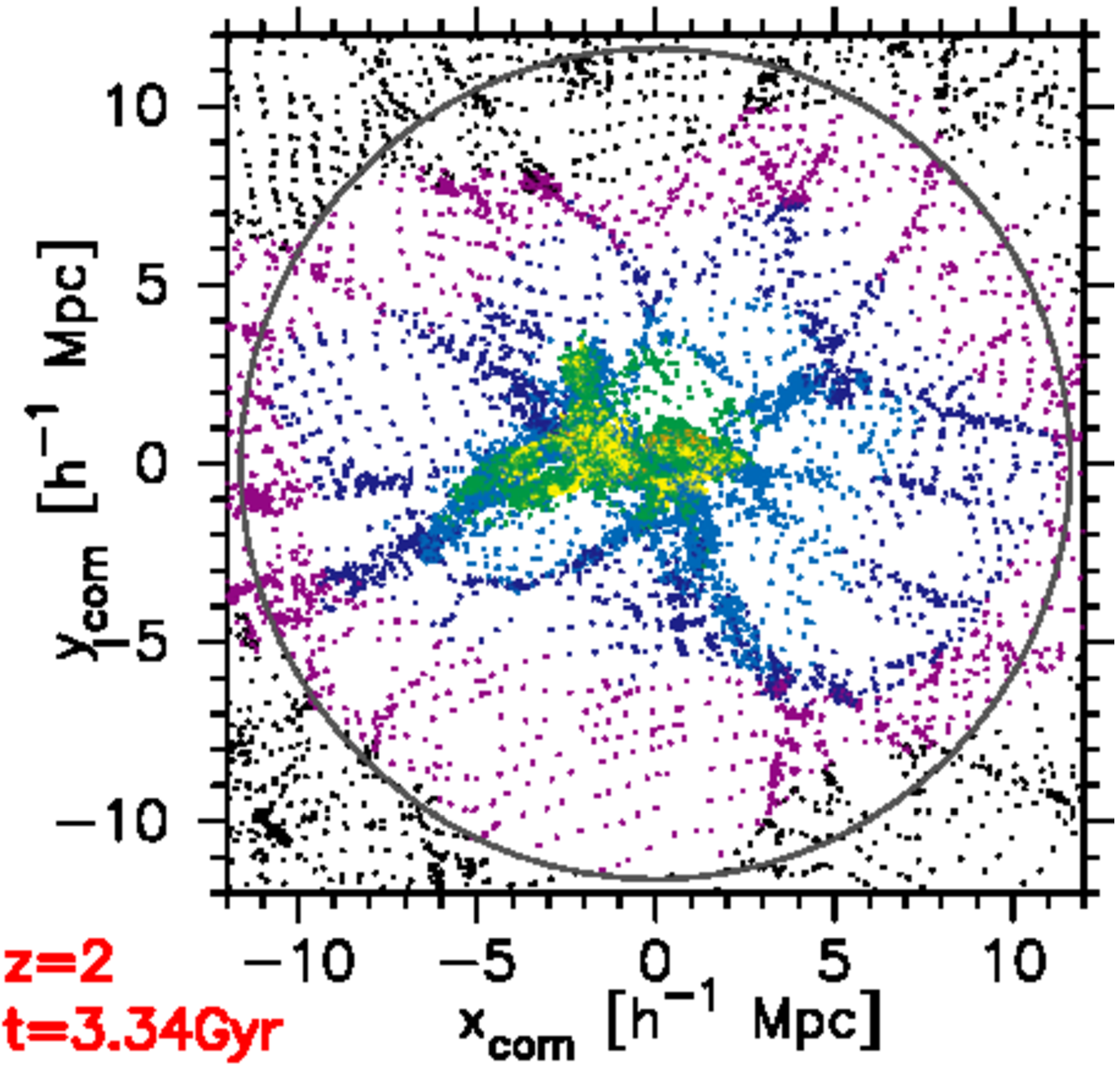}\qquad
\includegraphics[width=6cm]{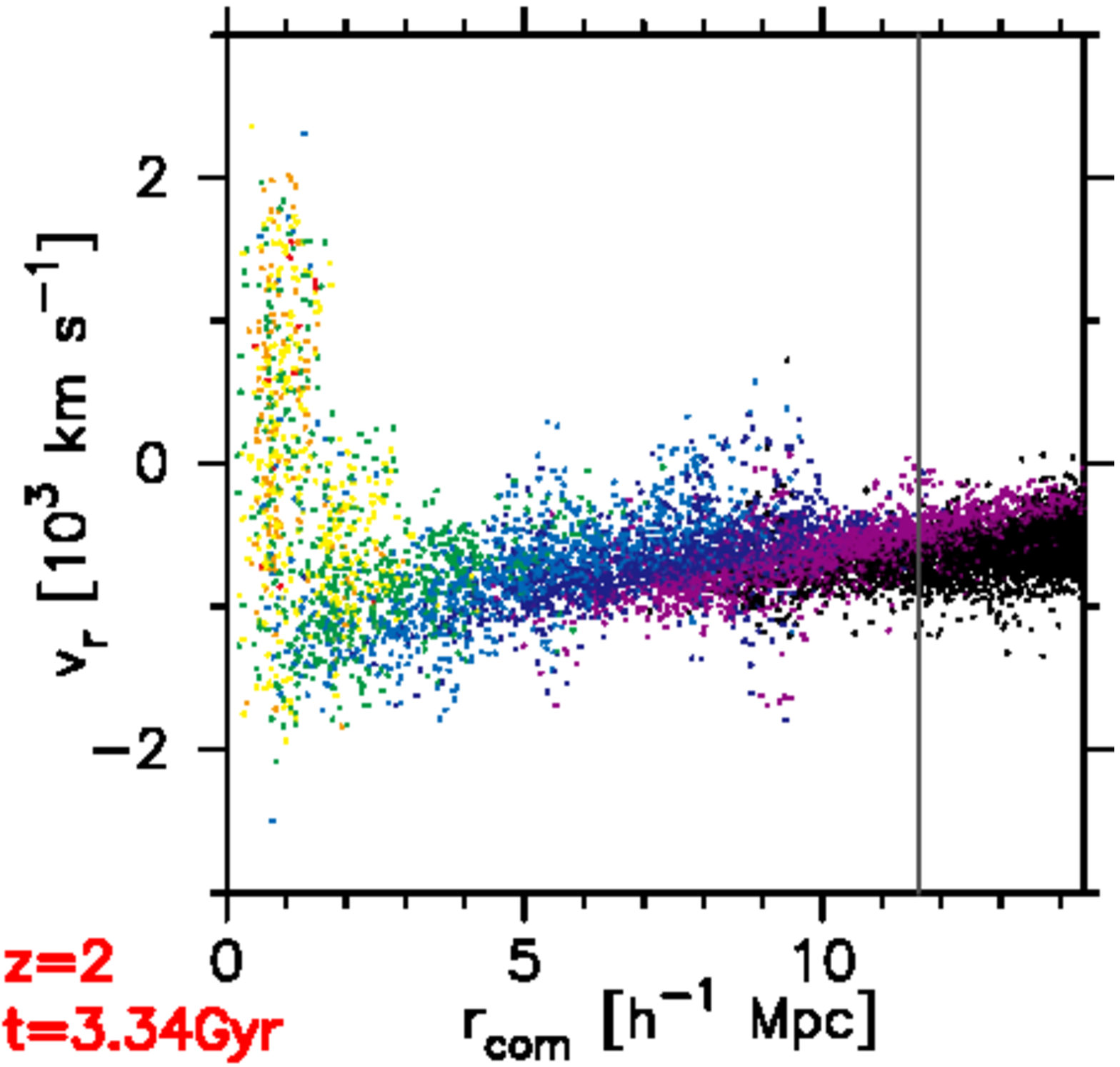}
\end{center}
\caption{The particle distribution of Halo I in the comoving space (left) and the phase space (right).  We select a slice of thickness 3 \% of $R_M(z)$ at each redshift, and plot all the particles within the slice in the lect panel. In contrast, we consider a large sphere that encloses the protohalo defined at each halo and plot  randomly selected 1 \% of the particles in the sphere. The gray circles in the left panels and the vertical lines in the right panels indicate $R_M(z)$. The particles are color-coded according to the initial position at $z=99$; the sphere of radius $R_M(z = 99)$ is divided into seven equal radial shells, and particles in each bin are plotted in different colors.  Black points correspond to particles outside the initial halo at $z = 99$. Those different color particles become mixed due to the subsequent evolution. In order to clarify the later evolution visually . we redefine the colors of the particles at $z = 1$ (approximately the turn-around epoch), and keep the color convention until $z=0$.}
\label{dist}
\end{figure*}

\addtocounter{figure}{-1}
\begin{figure*}[h]
\begin{center}
\includegraphics[width=6cm]{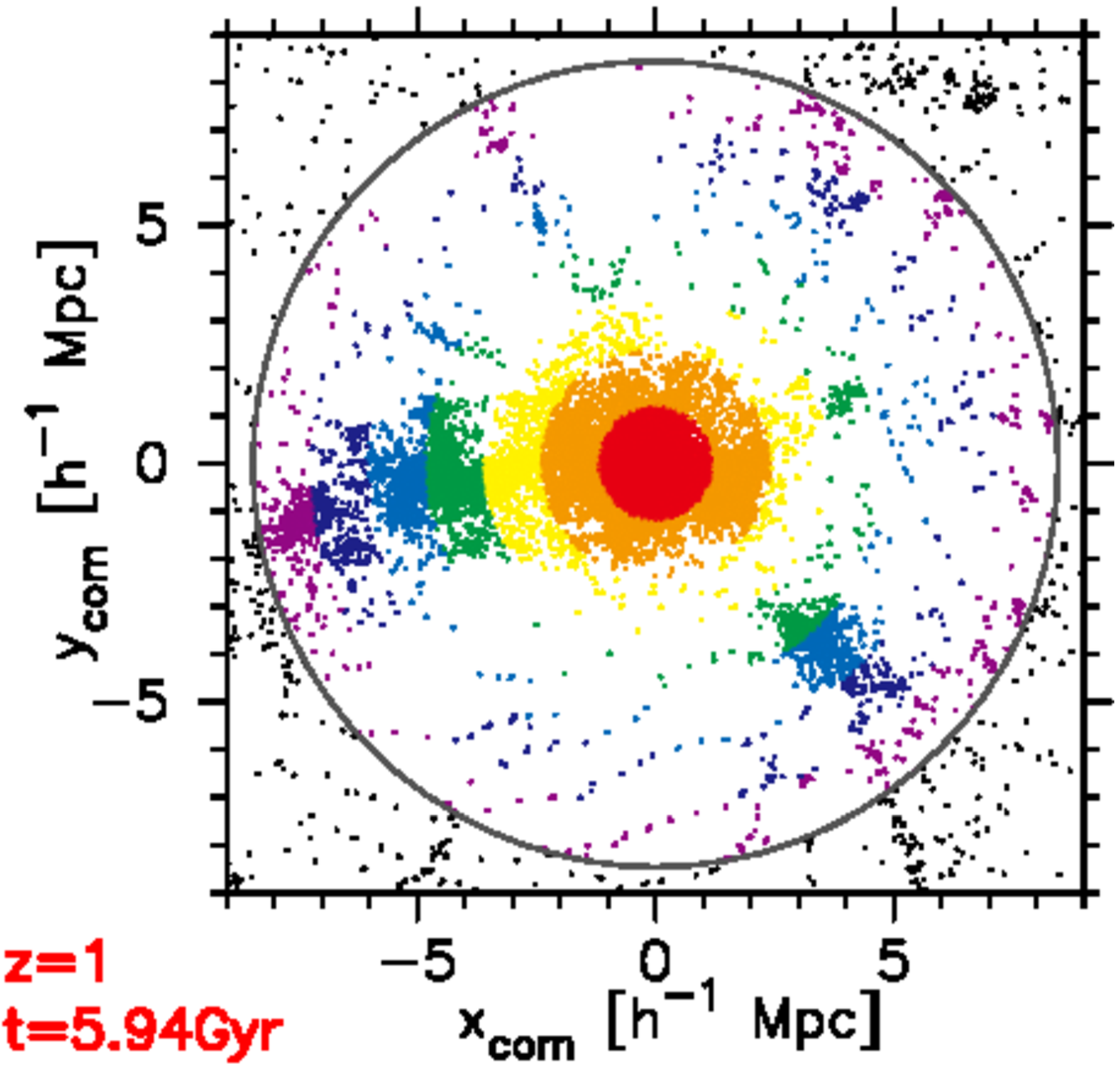}\qquad
\includegraphics[width=6cm]{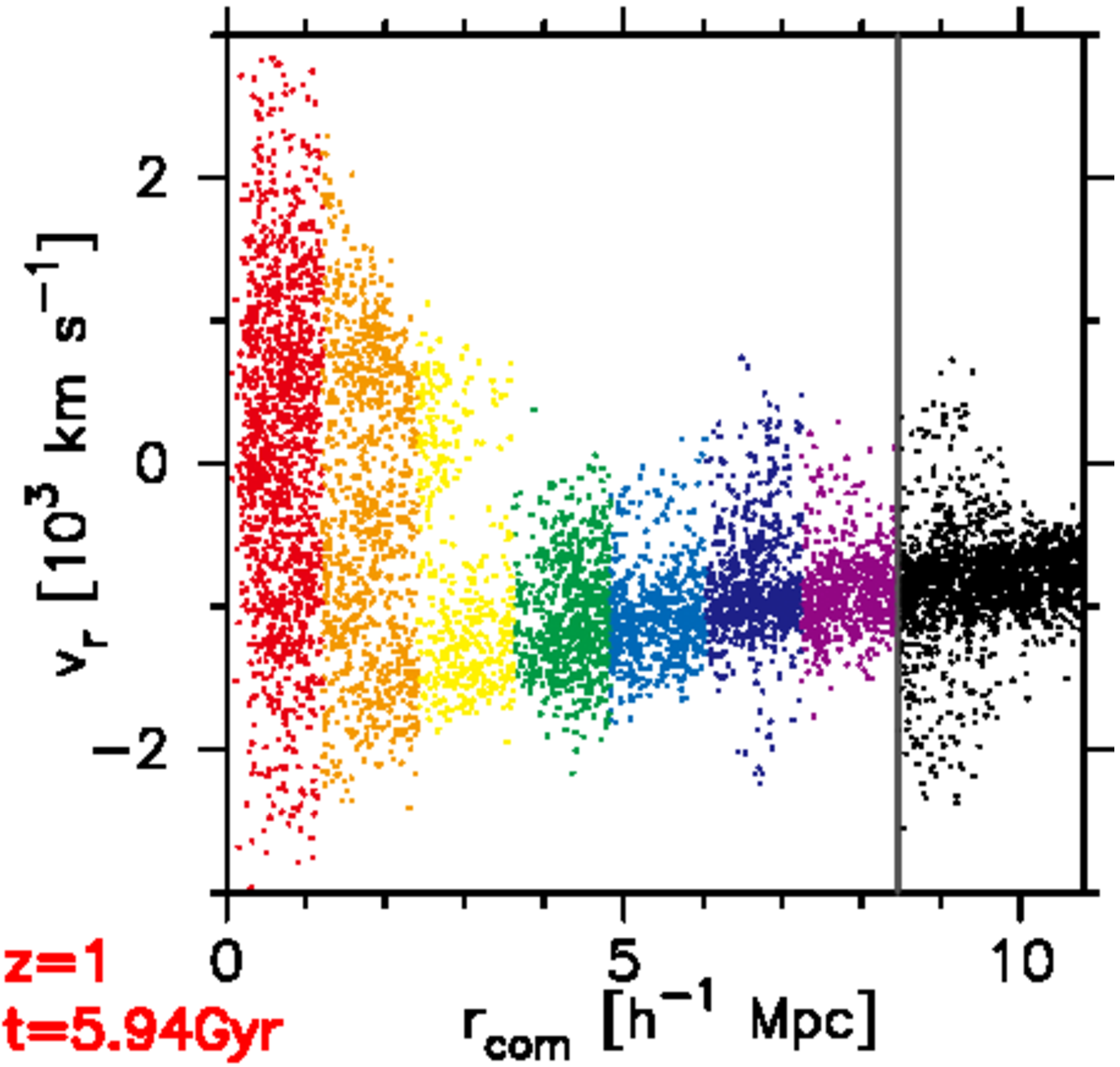}\\
\includegraphics[width=6cm]{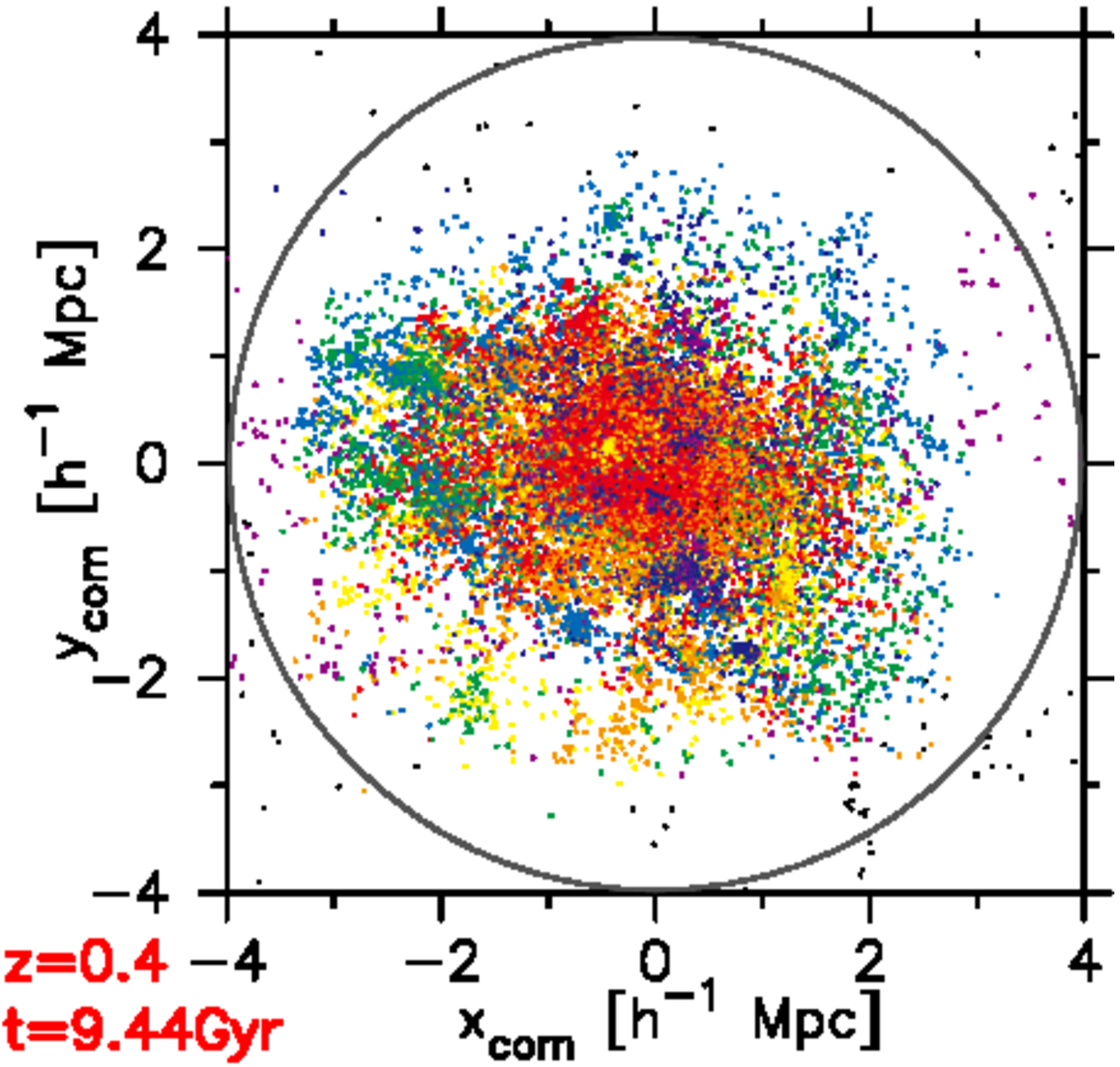}\qquad
\includegraphics[width=6cm]{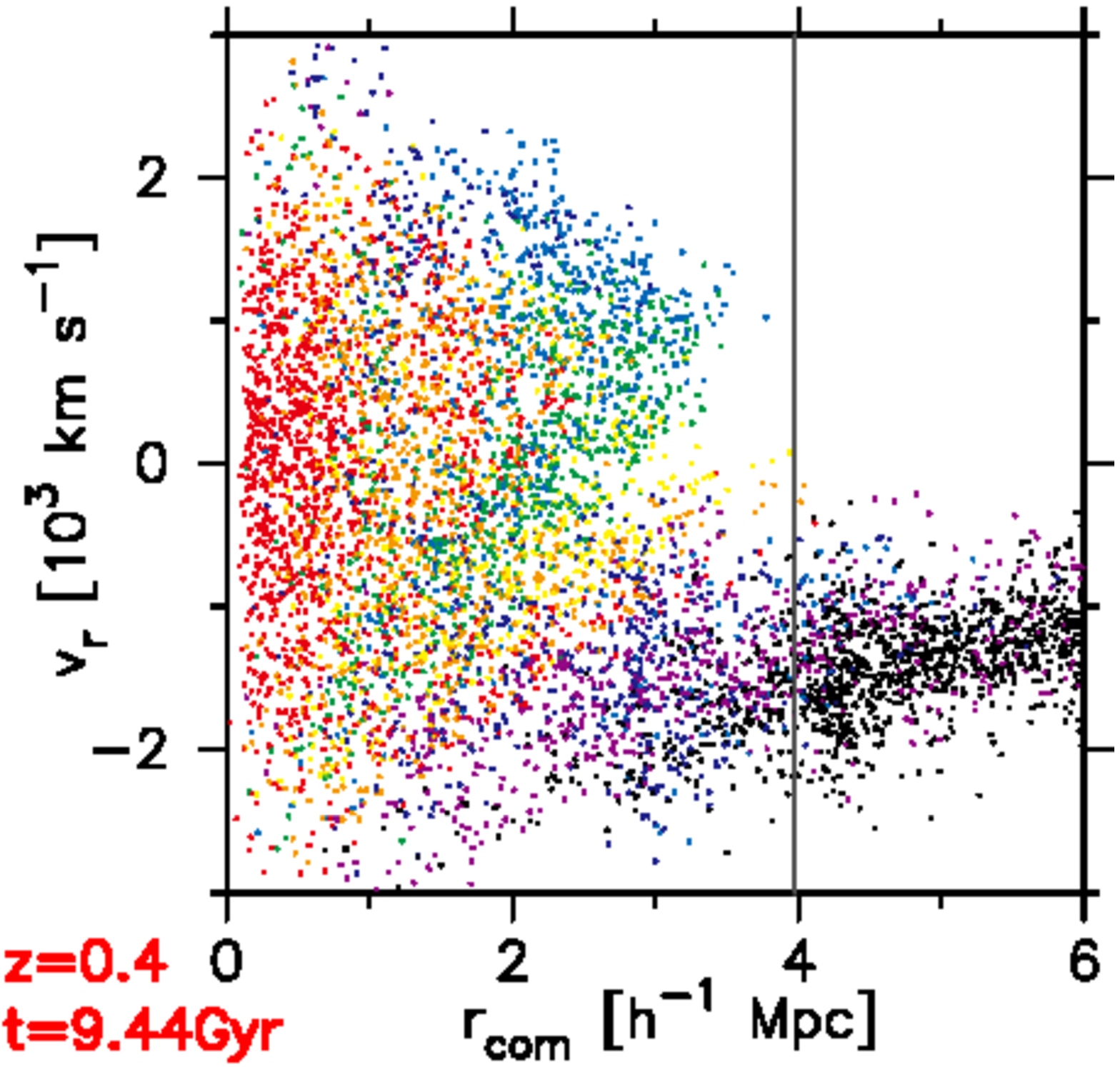}\\
\includegraphics[width=6cm]{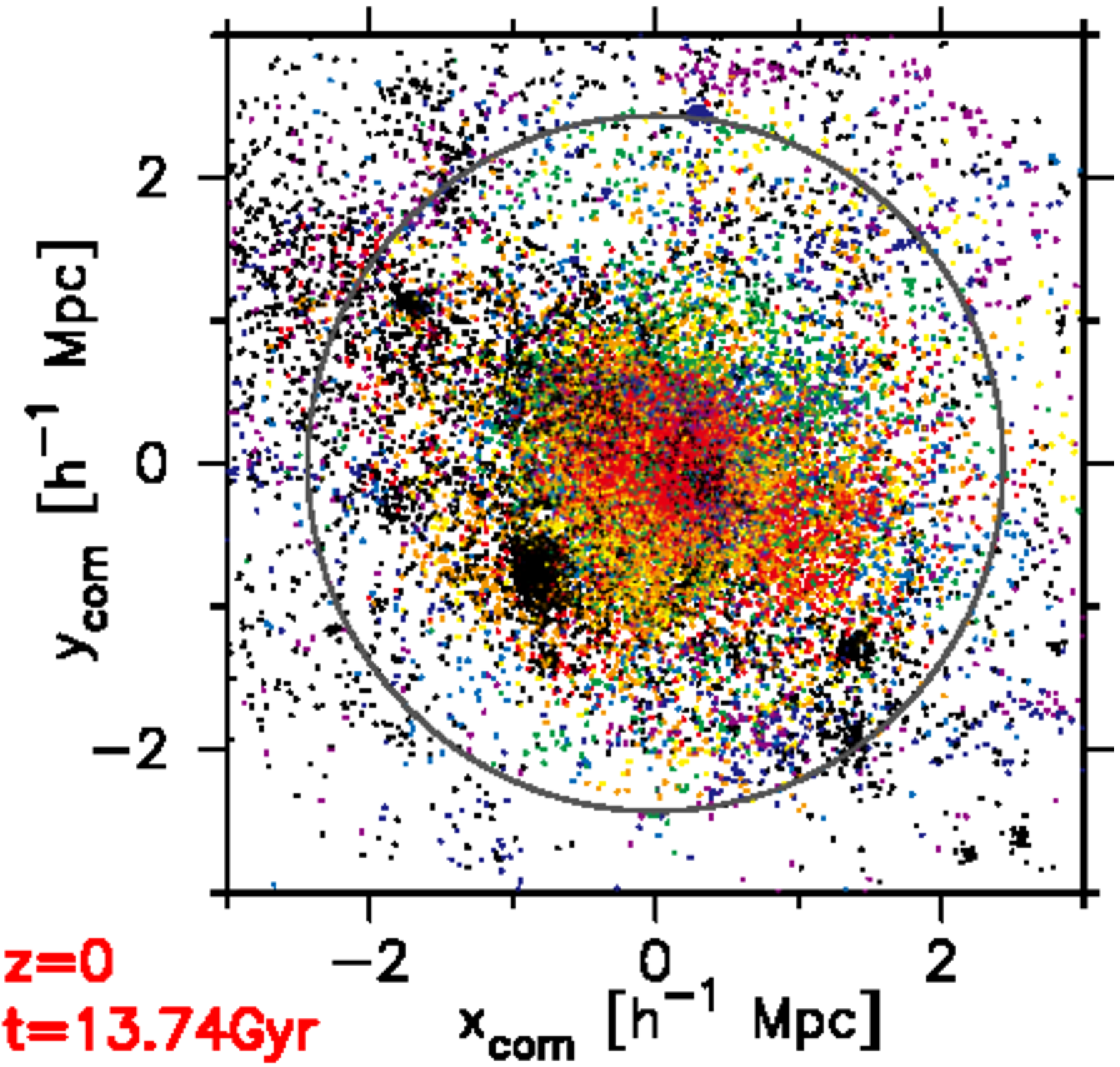}\qquad
\includegraphics[width=6cm]{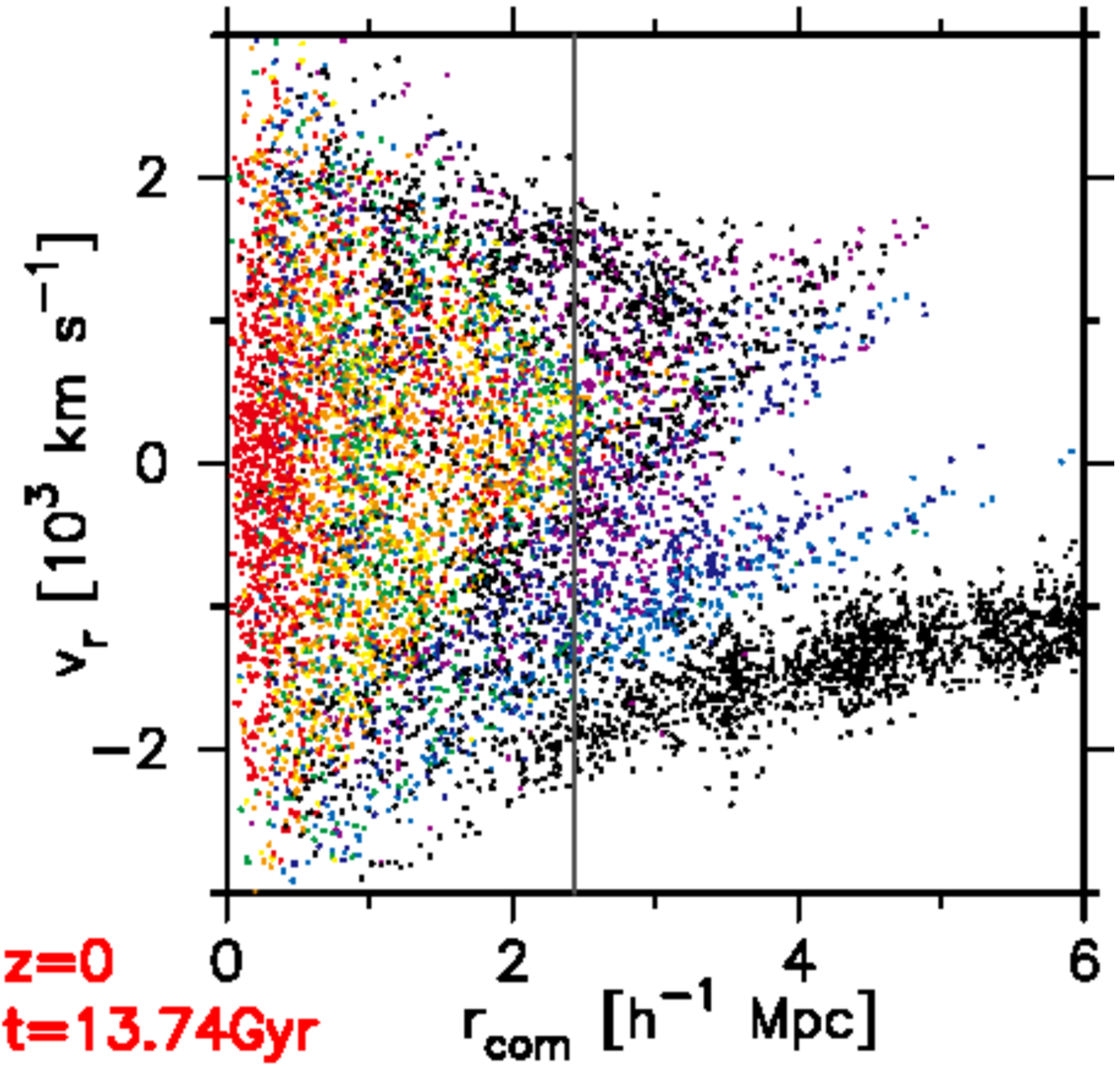}
\end{center}
\caption{Continued.}
\end{figure*}

\begin{figure*}[h]
\begin{center}
\includegraphics[width=6cm]{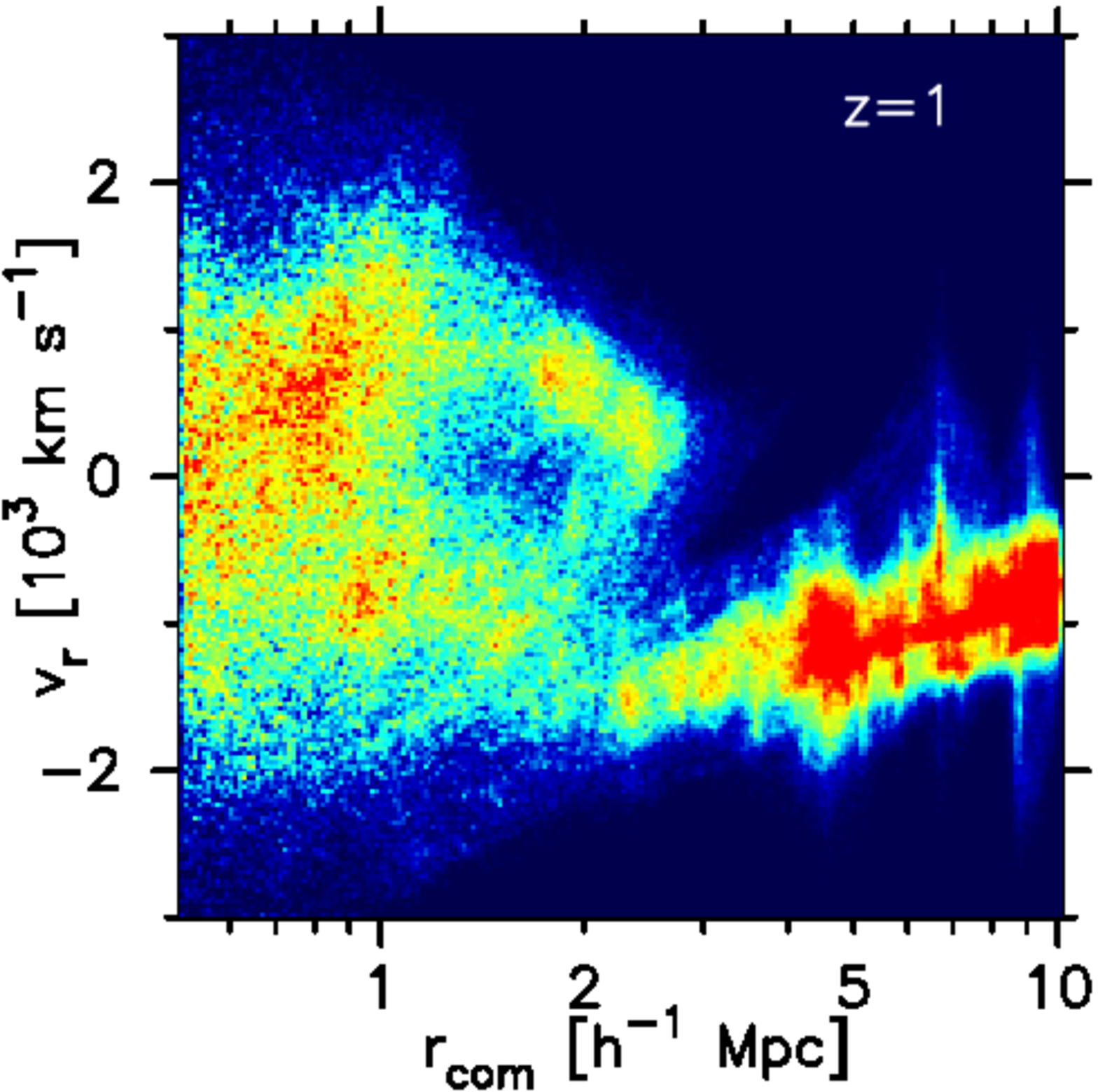}\qquad
\includegraphics[width=6cm]{vprof1.eps}\\
\includegraphics[width=6cm]{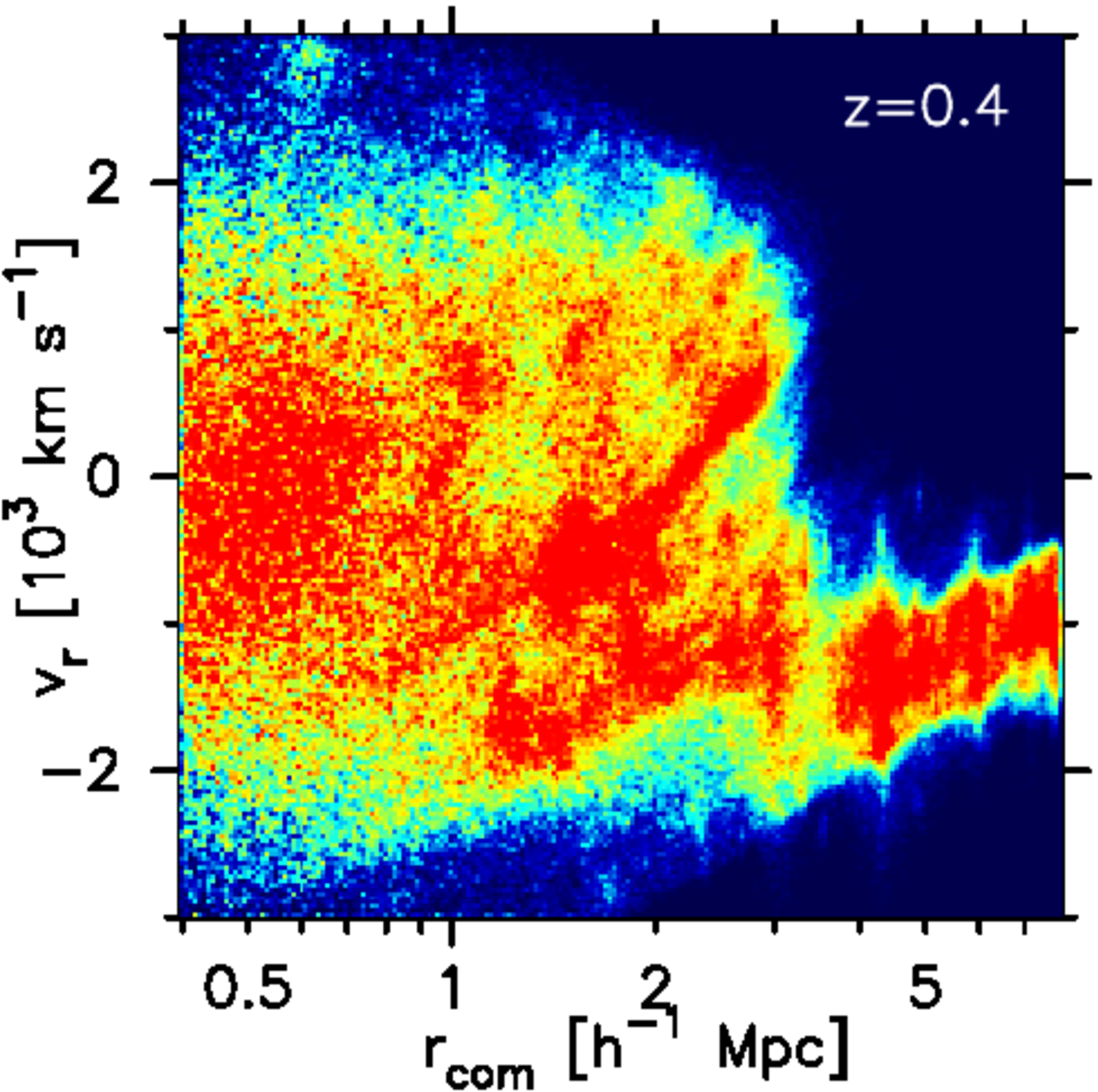}\qquad
\includegraphics[width=6cm]{vprof2.eps}\\
\includegraphics[width=6cm]{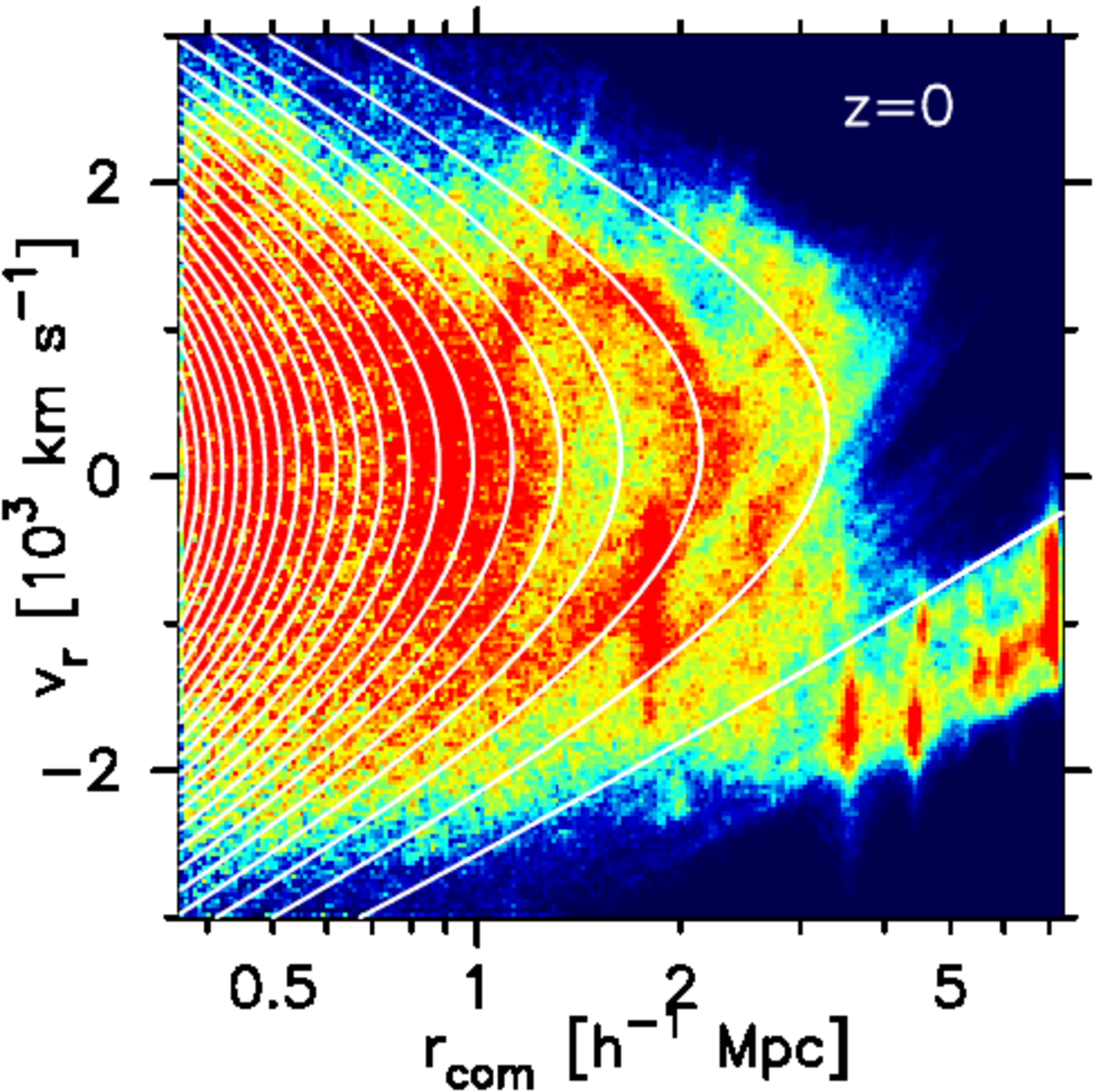}\qquad
\includegraphics[width=6cm]{vprof3.eps}
\end{center}
\caption{Halo I's color contrast images of the phase space density (left) and the corresponding profiles (right) of radial velocity (green dashed), radial (black solid) and tangential (red dotted) velocity dispersions, for $z=$1 (top), 0.4 (middle) and 0 (bottom). For the contrast image for $z=0$, the self-similar solution \citep{Filmore84,Bertschinger85} in the EdS universe is overplotted.}
\label{colcont}
\end{figure*}

\begin{figure*}[h]
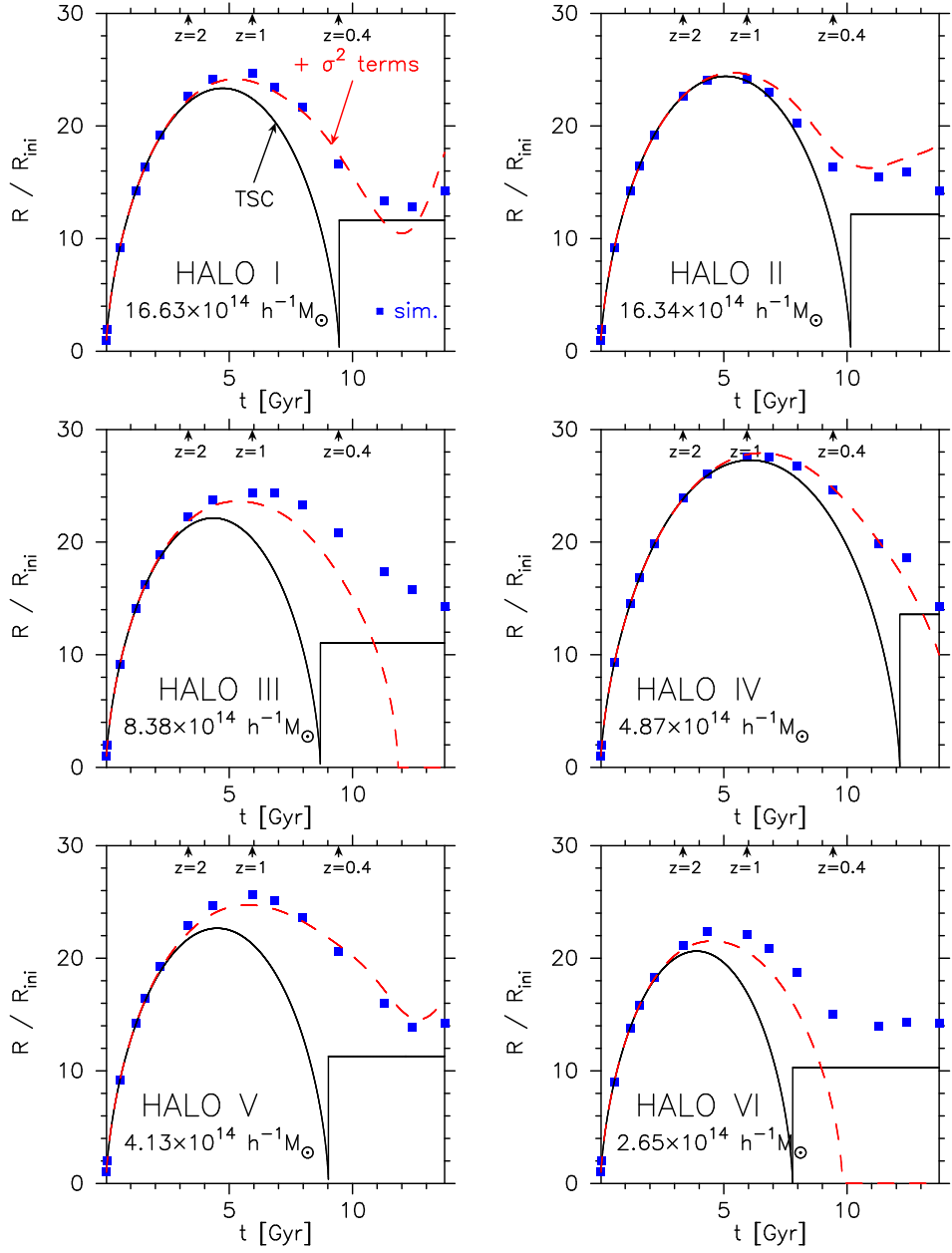

\begin{center}
\includegraphics[width=6cm]{collapse2.eps}\qquad
\includegraphics[width=6cm]{collapse1.eps}\\
\includegraphics[width=6cm]{collapse10.eps}\qquad
\includegraphics[width=6cm]{collapse45.eps}\\
\includegraphics[width=6cm]{collapse68.eps}\qquad
\includegraphics[width=6cm]{collapse80.eps}
\end{center}
\caption{The comparison of the evolution of the halo radius predicted by TSC (solid) with the simulation (squares). The model prediction is calculated by using the initial condition of each simulated halos. The red dashed line shows the solution of the collapse model with the velocity dispersion terms included (see text), which improves the prediction for the evolution of the halo radius.}
\label{collapse}
\end{figure*}

\begin{figure*}[h]
\begin{center}
\includegraphics[width=6cm]{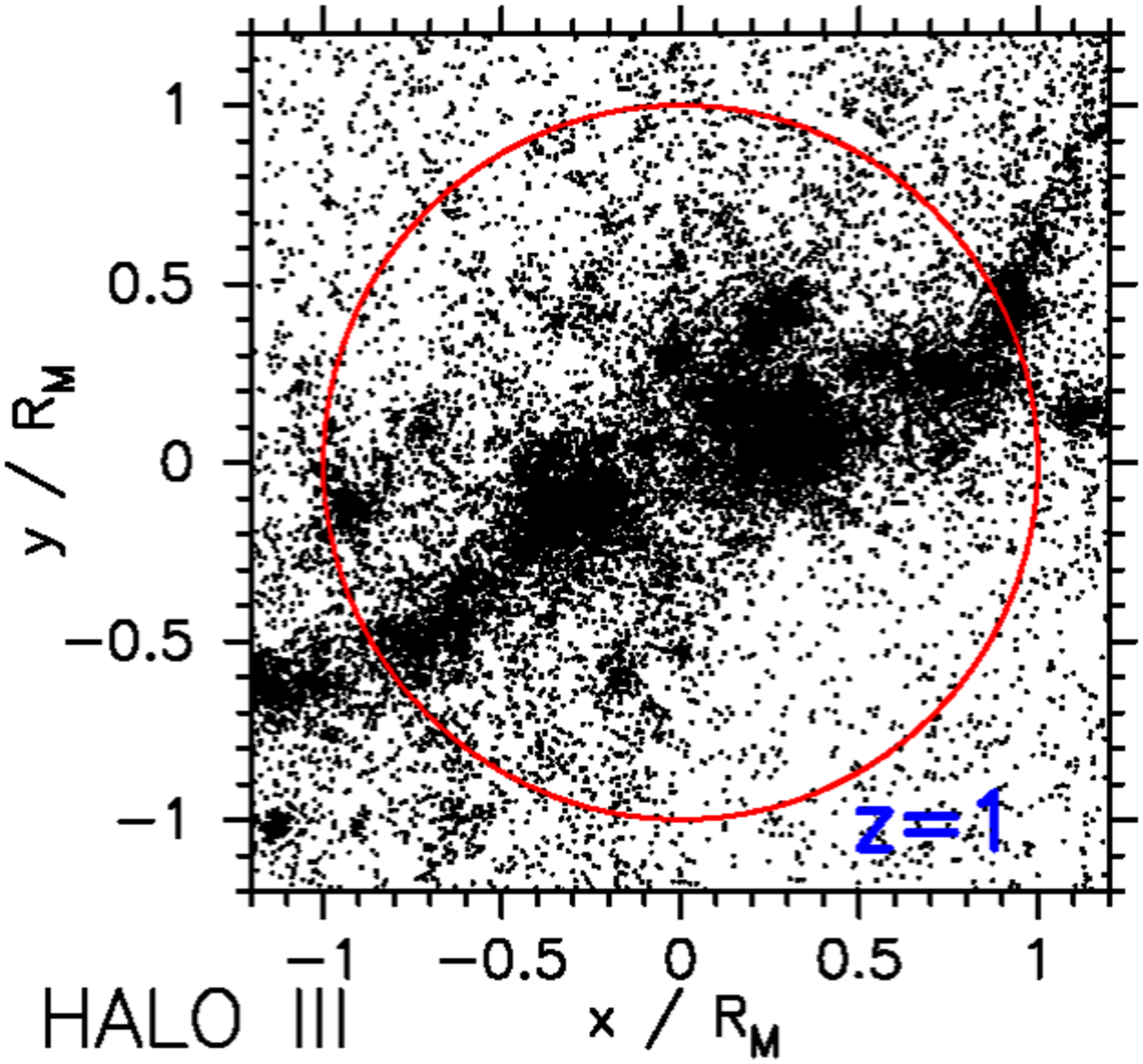}\qquad
\includegraphics[width=6cm]{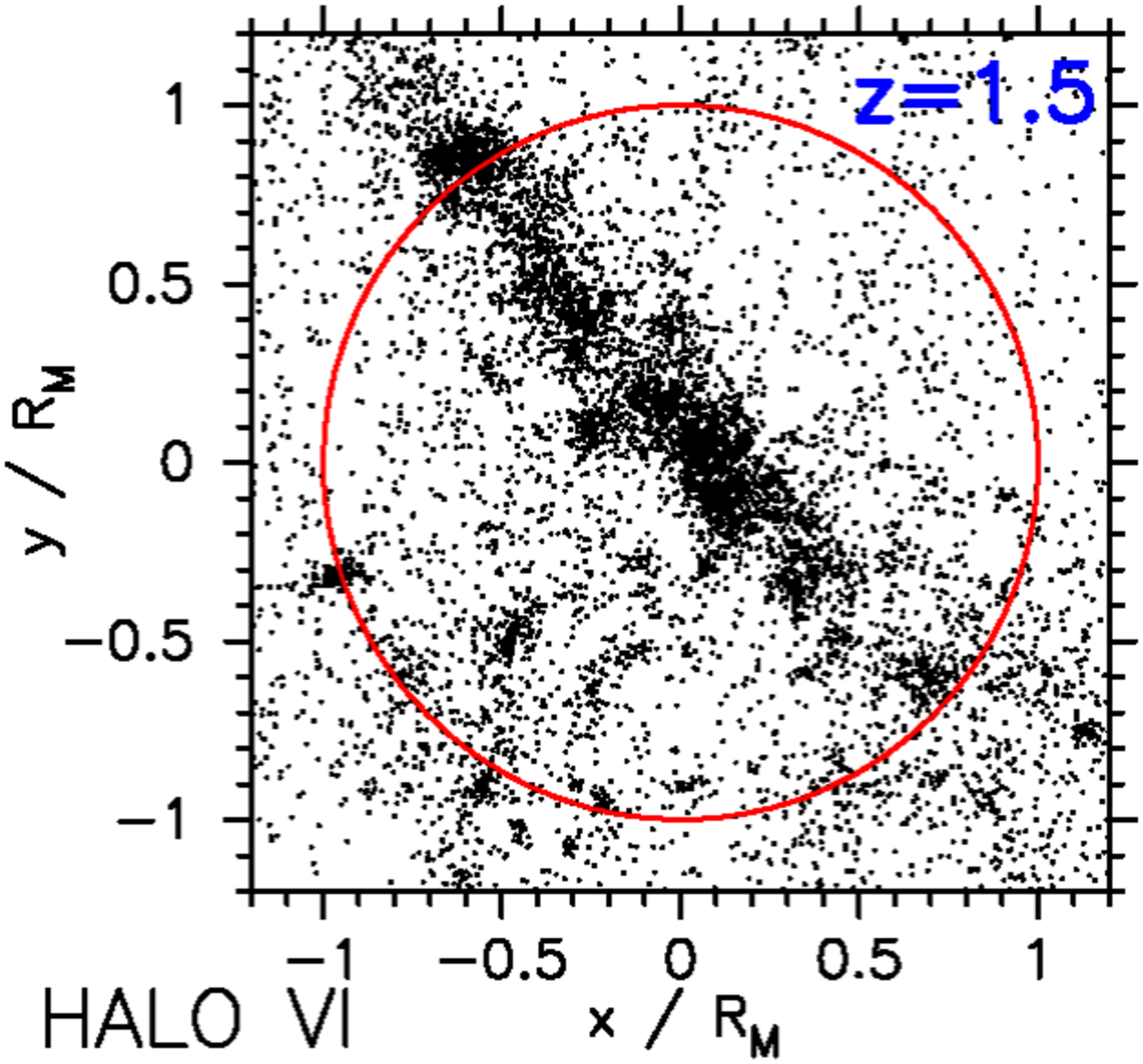}
\end{center}
\caption{The particle distributions of Halo III at $z=1$, and of Halo VI at $z=1.5$, showing their highly non-spherical evolution. The plotted particles are a randomly selected 5 \% of those in the box 1.2 $R_M(z)$ on a side, centered on the halo center. }
\label{dists}
\end{figure*}

\begin{figure*}[h]
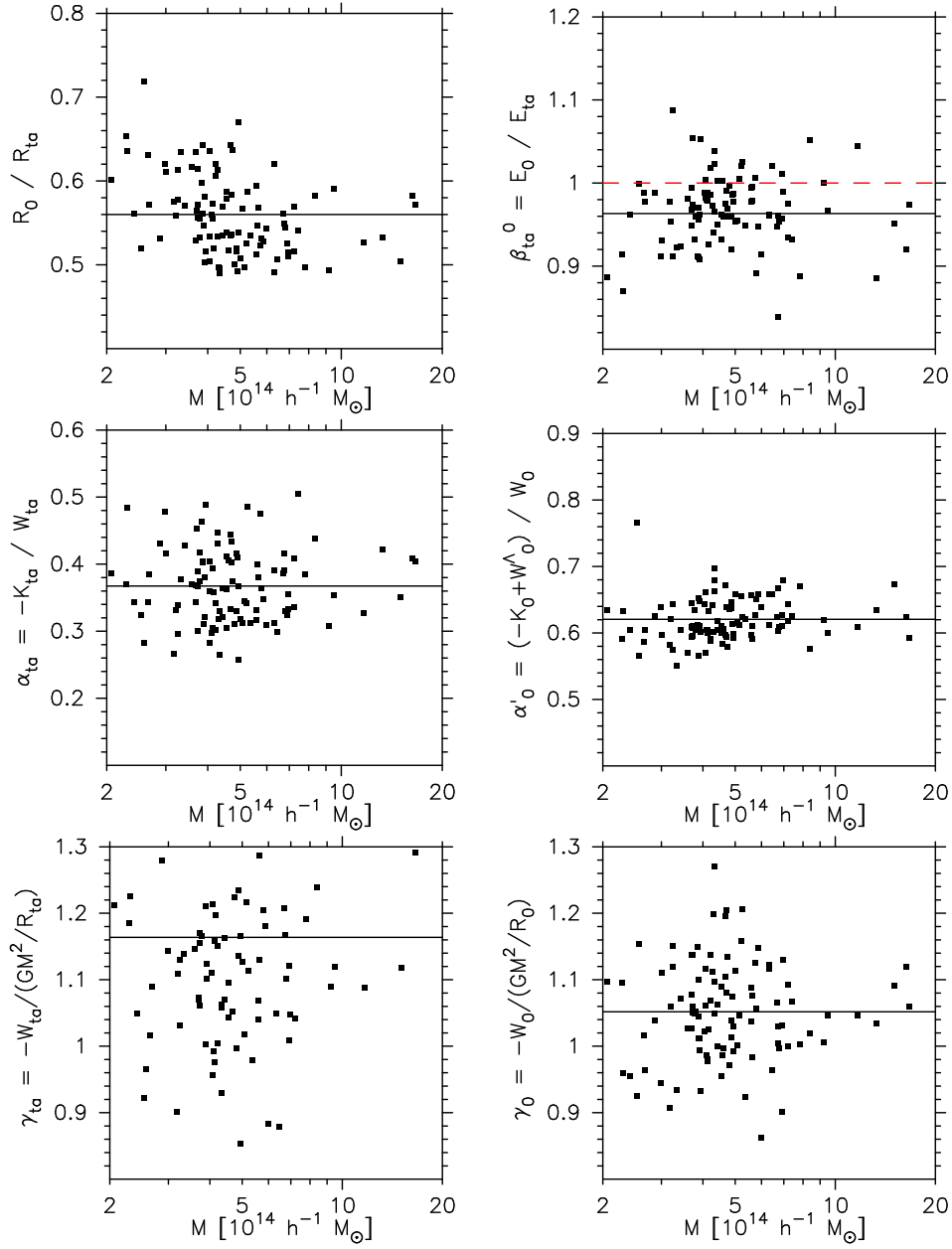

\begin{center}
\includegraphics[width=6cm]{r0torta.eps}\qquad
\includegraphics[width=6cm]{betata.eps}\\
\includegraphics[width=6cm]{alphata.eps}\qquad
\includegraphics[width=6cm]{alpha0.eps}\\
\includegraphics[width=6cm]{gammata.eps}\qquad
\includegraphics[width=6cm]{gamma0.eps}
\end{center}
\caption{Comparison of the turn-around radius with the present radius of
the 100 simulated halos (top-left). Their difference can be attributed
to the parameters $\alphata$ (middle-left), $\alpha'_0$ (middle-right),
$\gammata$ (bottom-left) and $\gamma_0$ (bottom-right). The energy
conservation is also checked in terms of $\betata$ (top-right). The
solid line in each panel indicates the average value. The red dashed line in the top-right panel shows unity
, meaning the total energy in the sphere is conserved between the two epochs.}
\label{tato0}
\end{figure*}

\begin{figure*}[h]
\begin{center}
\includegraphics[width=6cm]{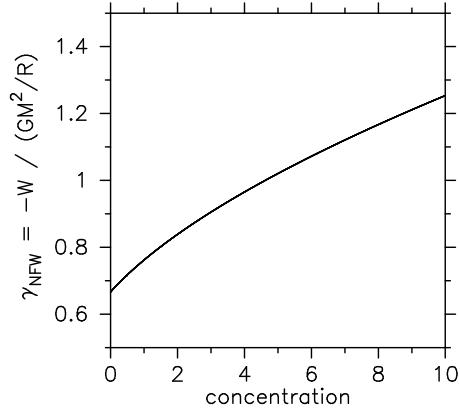}
\end{center}
\caption{The parameter $\gamma$, defined by $W=-\gamma GM^2/R$, for the
NFW density profile at the virial radius as a function of the
concentration parameter.}  \label{gammaNFW}
\end{figure*}

\begin{figure*}[h]
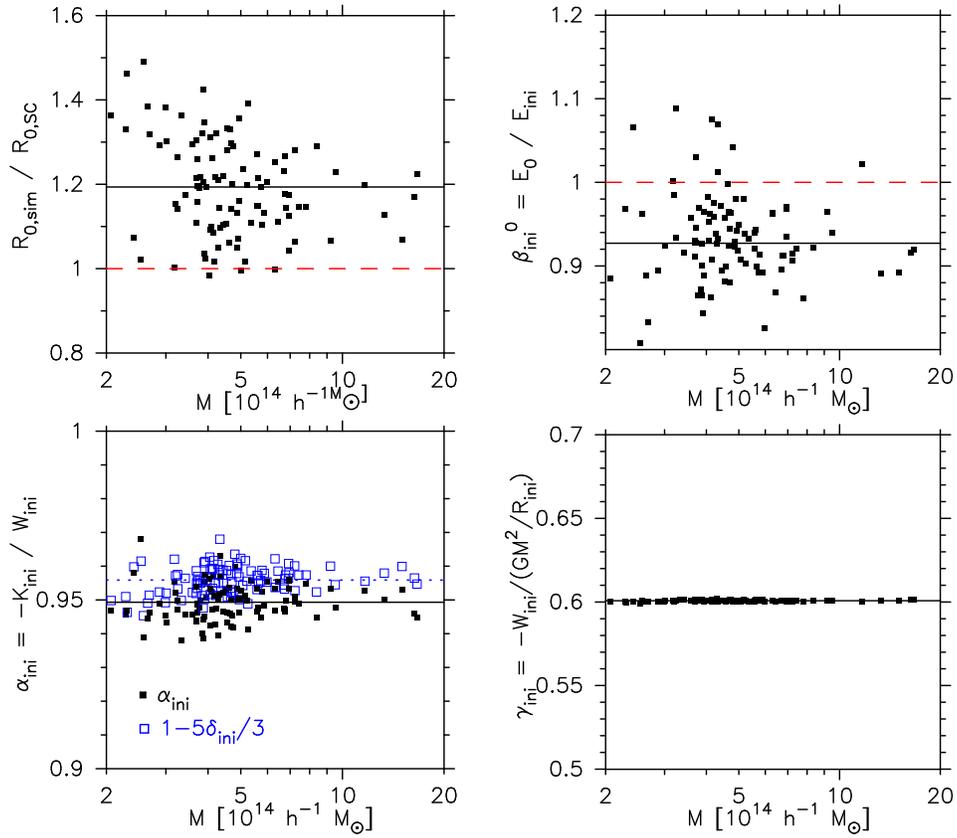

\begin{center}
\includegraphics[width=6cm]{r0simtor0mod.eps}\qquad
\includegraphics[width=6cm]{betai.eps}\\
\includegraphics[width=6cm]{alphai.eps}\qquad
\includegraphics[width=6cm]{gammai.eps}\\
\end{center}
\caption{Comparison of the prediction of TSC with the present radius of the simulated halos (upper-left). Their difference is attributed to $\alphai$ (lower-left), $\betai$ (upper-right) and $\gammai$ (lower-right), and $\alpha'_0$ and $\gamma_0$ in Figure \ref{tato0} (see text for the definition of the parameters). The solid line in each panel indicates the average value. The parameter $\alphai$ is compared with the theoretical prediction 1-5$\deltai$/3 (blue open squares), and its average value is shown by the blue dotted line. The red dashed lines in the upper panels show unity for comparison; if TSC prediction is perfect, the radios of the radii and total energies become unity.}
\label{inito0}
\end{figure*}

\begin{figure*}[h]
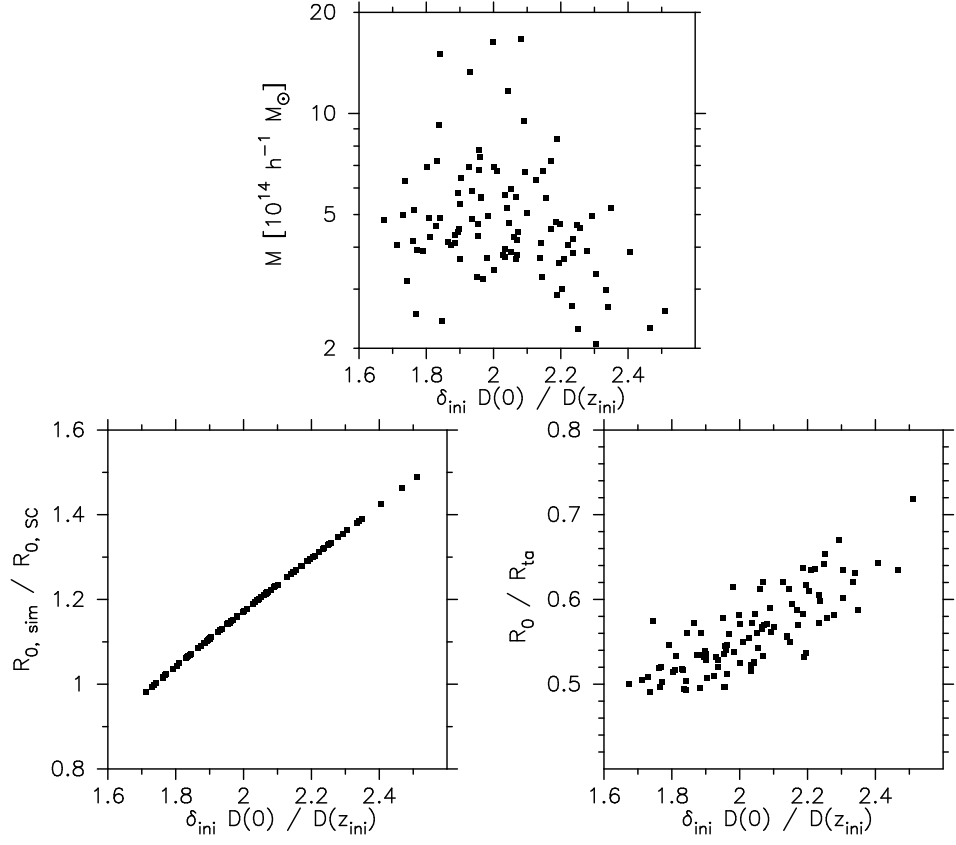

\begin{center}
\includegraphics[width=6cm]{deliniM.eps}\\
\includegraphics[width=6cm]{deliniR0R0.eps}\qquad
\includegraphics[width=6cm]{deliniR0Rta.eps}
\end{center}
\caption{The halo mass (upper) and the difference in the radial ratios between TSC and the simulation (lower), vs. the initial overdensity (normalized by the linear growth factor).}
\label{delini}
\end{figure*}

\end{document}